\documentclass{aa}
\usepackage{amssymb}
\usepackage{graphicx}
\usepackage{hyperref}

\def\asec{\ifmmode ^{\prime\prime}\else$^{\prime\prime}$\fi}

\def\degs{\ifmmode ^{\circ}\else$^{\circ}$\fi}
\def\amin{\ifmmode ^{\prime}\else$^{\prime}$\fi}
\def\asec{\ifmmode ^{\prime\prime}\else$^{\prime\prime}$\fi}
\def\fss{\hbox{$.\!\!^{\rm s}$}}        % Fractions of seconds
\def\fdg{\hbox{$.\!\!^\circ$}}          % Fractions of degrees
\def\farcs{\hbox{$.\!\!^{\prime\prime}$}}  % Fractions of arcseconds
\def\h{$^{\rm h}$}
\def\m{$^{\rm m}$}

\def\degs{\ifmmode ^{\circ}\else$^{\circ}$\fi}
\def\amin{\ifmmode ^{\prime}\else$^{\prime}$\fi}

\def\widerul{\vrule height 2.5ex width 0ex depth 0ex}
\def\wideru{ \vrule height 2.8ex   width 0ex depth 0ex}
\def\widerul{\vrule height 2.5ex width 0ex depth 0ex}

\def\twolines#1#2{
\renewcommand{\arraystretch}{0.8}
\begin{tabular}{@{}c@{}}
#1 \vrule height3.2ex width0ex \\ #2 \\[.7ex]
\end{tabular}
}

\begin{document}

%\thesaurus{08.19.5 SN 1987A; 02.14.1; 08.19.4} 

\unitlength=1mm
\def\EE#1{\times 10^{#1}}
\def\gcm{\rm ~g~cm^{-3}}
\def\cm3{\rm ~cm^{-3}}
\def\kms{\rm ~km~s^{-1}}
\def\cms{\rm ~cm~s^{-1}}
\def\ergs{\rm ~erg~s^{-1}}
\def\wl{~\lambda}
\def\wll{~\lambda\lambda}
\def\Nii{M(^{56}{\rm Ni})}
\def\FeI{{\rm Fe\,I}}
\def\FeII{{\rm Fe\,II}}
\def\FeIII{{\rm Fe\,III}}
\def\Niii{M(^{57}{\rm Ni})}
\def\FeIb{{\rm [Fe\,I]}}
\def\FeIIb{{\rm [Fe\,II]}}
\def\FeIIIb{{\rm [Fe\,III]}}
\def\OIb{{\rm [O\,I]}}
\def\OIIb{{\rm [O\,II]}}
\def\OIIIb{{\rm [O\,III]}}
\def\SIIb{{\rm [S\,II]}}
\def\ArIIIb{{\rm [Ar\,III]}}
\def\NiIIb{{\rm [Ni\,II]}}
\def\Msun{~{\rm M}_\odot}
\def\Ti44{M(^{44}{\rm Ti})}
\def\MZA{M_{\rm ZAMS}}
\def\mum{\mu{\rm m}}
\def\muJ{\mu{\rm Jy}}
\def\psr{PSR~B0540-69.3}
\def\snr{SNR~0540-69.3}

\def\lsim{\!\!\!\phantom{\le}\smash{\buildrel{}\over
  {\lower2.5dd\hbox{$\buildrel{\lower2dd\hbox{$\displaystyle<$}}\over
                               \sim$}}}\,\,}
\def\gsim{\!\!\!\phantom{\ge}\smash{\buildrel{}\over
  {\lower2.5dd\hbox{$\buildrel{\lower2dd\hbox{$\displaystyle>$}}\over
                               \sim$}}}\,\,}

\title{The young pulsar PSR~B0540-69.3 and its synchrotron nebula in 
the optical and X-rays\thanks{Based on observations performed at the European Southern 
Observatory, Paranal, Chile (ESO Program 67.D-0519).}}

\author {N.~I.~Serafimovich\inst{1,2}
\and Yu.~A.~Shibanov\inst{1}
\and P.~Lundqvist\inst{2} 
\and J.~Sollerman\inst{2} 
}

\institute{Ioffe Physical Technical Institute, Politekhnicheskaya 26, 
St. Petersburg, 194021, Russia
\and Stockholm Observatory, AlbaNova Science Center, Department 
of Astronomy, SE-106 91 Stockholm, Sweden
}

%\date{Submitted}

\date{Received 22 March 2004; accepted 22 June 2004}
%\mail{peter@astro.su.se}

\titlerunning{PSR B0540-69.3 and its PWN}
\authorrunning{Serafimovich et al.}
\offprints{N.~Serafimovich;\hfill\\
e-mail: natalia@astro.su.se}

\abstract{
The young PSR B0540-69.3 in the LMC is the only pulsar (except the Crab 
pulsar) for which a near-UV spectrum has been obtained. 
However, the absolute flux and spectral index of the HST/FOS spectrum are 
significantly higher than suggested by previous broad-band time-resolved 
groundbased UBVRI photometry. To investigate this difference,
observations with ESO/VLT/FORS1 and analysis of HST/WFPC2 archival data 
were done. We show that the HST and VLT spectral data for the pulsar 
have $\gsim 50$\% nebular contamination and that this is the reason for 
the above mentioned difference. The broadband HST spectrum 
for the range 3300--8000~\AA\ is clearly nonthermal and has a negative 
spectral index, $F_{\nu} \propto \nu^{-\alpha}$ 
with $\alpha_{\nu} = 1.07^{+0.20}_{-0.19}$. 
This is different from the almost flat spectrum of the Crab pulsar, and 
also steeper than for the previously published broadband photometry of \psr. 
We have also studied the spatial
variations of the brightness and spectral index of the Pulsar Wind Nebula 
(PWN) around the pulsar, and find no significant spectral index variation
over the PWN. The HST data show a clear asymmetry of the surface brightness 
distribution along the major axis of the torus-like structure of the PWN with 
respect to the pulsar position, also seen in Chandra/HRC X-ray images.
This is different from the Crab PWN and likely 
linked to the asymmetry of the surrounding SN ejecta.  
The HST/WFPC2 archival data have an epoch separation by 4 years, and this
allows us to estimate the proper motion of the pulsar. We find a motion 
of $4.9\pm2.3$ mas yr$^{-1}$ (corresponding to a transverse velocity
of $1190\pm560 \kms$) along the southern jet of the PWN.  
If this is confirmed at a higher significance level 
by future observations, this makes \psr\ 
the third pulsar with 
a proper motion aligned with the jet axis of its PWN, which poses 
constraints on pulsar kick models. To establish the multiwavelength 
spectrum of the pulsar and its PWN, we have included recent Chandra 
X-ray data, and discuss the soft pulsar X-ray spectrum based on spectral fits 
including absorption by interstellar gas in the Milky Way, LMC as well as the 
supernova ejecta. We have compared the multiwavelength spectra of \psr\ and
the Crab pulsar, and find that both \psr\ and the Crab pulsar have a weaker
flux in the optical than suggested by a low-energy power-law extension of
the X-ray spectrum. This optical depression is more severe for \psr\ than for
the Crab pulsar. The same trend is seen for the PWNe of the two pulsars,
and continues for low energies also out in the radio band. We discuss 
possible interpretations of this behavior.
}

\maketitle

\section{Introduction}

\psr~in the Large Magellanic Cloud (LMC) was discovered as a pulsed 
($P = 50.2$~ms) X-ray source by Seward et al. (1984). Pulsations 
have since also been detected in the optical and at~radio 
wavelengths (\cite{MP85}; \cite{Man93a}).
The pulse profile in the optical (\cite{Boyd95}) is broad and double-peaked, 
with a separation of $\sim$~0.2
in phase between the two maxima, consistent with what is also seen in 
X-rays (\cite{SHH84}; \cite{Plaa03}).
The profile is also broad in the radio (the duty cycle is $\gsim$ 80\%), 
and there is a hint of a double structure (\cite{Man93a}).

Parameters of \psr~are compiled in Table~1. The pulsar
spins rapidly, is young (spin down age 1660 yr), and sits in a compact
synchrotron nebula (see Fig.~\ref{f:F547im}), which we will henceforth refer 
to as its pulsar wind nebula (PWN). The similarities with the Crab
pulsar and its nebula are 
such that \psr~with its supernova remnant, SNR~0540-69.3, are 
sometimes referred to as the ``Crab twin''. Even the structures of the PWNe 
appear to be similar. Both have a torus and jets 
(\cite{GW00}), although the  proper 
motion for \psr, suggested by Manchester et al. (1993b) based on a 
displacement between 
the pulsar optical  position and the center of the PWN as seen in radio, 
seems not to be along the spin axis as it is in the Crab  case.    
%-----------------------table1------------------------------
%-----------------------table1------------------------------
\begin{table*}[hbt]
\caption{Parameters of \psr\ (\cite{Man93a}, unless specified otherwise).}
\label{t:param}
\begin{tabular}{ccccccccccc}
\hline\hline
\multicolumn{6}{c}{Observed}& &\multicolumn{4}{c}{Derived} \wideru \\
\cline{1-6}\cline{8-11}
$P$ & $\dot P$ & $\mu^a$ & $ $ 
& $l$, $b^b$ & $D\!M^c$ & & $\tau^d$ & $B^{e}$ & $\dot E$ & $d^{f}$ \wideru \\
ms & $10^{-15}$ & mas yr$^{-1}$ &  
& & cm$^{-3}$ pc & & yr & G  & erg s$^{-1}$ & kpc \\ %\widerul \\ 
\hline 
50.2 &
{$479.04\pm0.10$} & $\sim 5$ &  &
%\twolines{$-37.2\pm1.2$}{$28.2\pm1.3$} & $3.4\pm0.7$ &
\twolines{279\fdg7}{$-$31\fdg5} &
$146\pm4$ &&
1660&
$4.96 \times 10^{12}$ &
$1.495 \times 10^{38}$ &
$51\pm1.3$  \widerul\\
\hline
\end{tabular}
\begin{tabular}{ll}
$^a$~Indirect proper motion estimates (\cite{Man93b}) & $^d$~Spin-down age $P/2\dot P$ %(\cite{SHH84}) 
\\  %\wideru\\
$^b$~Galactic coordinates (\cite{Kaa01}) & $^e$~Magnetic field  for a 10 km radius NS  $3.2\times 10^{19}(P\dot
P)^{1/2}$ %(\cite{Got03})\\  %\wideru 
\\ 
$^c$~Dispersion measure & $^f$~Distance to the LMC (\cite{P03}) \\ % \wideru \\
\end{tabular}
\end{table*}

There are, however, differences on a larger scale. While the PWN of \psr~is 
surrounded by an X-ray and radio emitting outer shell of 
radius $\sim$30\arcsec, or $\sim$7.3~pc % at 50~kpc distance,    
(\cite{Man93b}; \cite{GW00}), an 
outer shell around the Crab is still not confirmed (although high-velocity 
gas has been revealed in the UV, \cite{SLL00}). 
Another difference is that SNR~0540-69.3 is oxygen-rich (e.g., \cite{Kir89};
Serafimovich et al. 2004), whereas the Crab Nebula has nearly 
normal solar abundances of metals (\cite{Bl92}~and references therein). 
It is therefore believed that the progenitor 
to \psr~was a much more massive star than the Crab progenitor (\cite{Kir89}).

\psr~is one of few pulsars for which a near-UV or optical spectrum has been
reported. Hill et al. (1997) obtained a time-integrated near-UV spectrum 
with HST/FOS and Middleditch \& Pennypacker (1985) used time-resolved
photometry to establish a broadband ground-based UBVRI spectrum in the 
optical. These two spectra show,
however, a significant difference in absolute flux in the spectral range 
where they overlap. To investigate this mismatch we have added two recent sets of
data, one is the ESO/VLT/FORS spectroscopy of SNR~0540-69.3 analyzed by 
Serafimovich et al. (2004), and the other is 
HST/WFPC2 imaging (Caraveo et al. 2000; \cite{Mo03}) retrieved from the HST
archive.
A bonus of our study is that we also obtain the optical 
spectrum of the PWN around \psr. This was first studied quantitatively by 
Chanan et al. (1984), albeit at a low spatial 
resolution which did not allow them to resolve the pulsar from the PWN.

To connect the optical pulsar emission to the emission at other wavelengths,
we have also included recent results from radio and X-rays. Previous attempts
to establish the multiwavelength spectrum of \psr\ have assumed a rather
high hydrogen column density for the X-ray 
absorption, $N_{\rm H} \sim 4.6\times10^{21}$~cm$^{-2}$ (Kaaret et al. 2001).
With this value for $N_{\rm H}$ it is possible to fit the
soft X-ray spectrum with a single power-law. 
%%%% 
This suggests a non-thermal nature of 
the emission, likely to be formed in the magnetosphere 
of the rotating neutron star (NS).
%%%%%%%%%
There are, however, reasons to reinvestigate this since the spectral
fits have not considered the fact that a large fraction of the absorbing gas 
has LMC abundances rather than Milky Way abundances.
It could even be that the supernova ejecta can contribute to the
absorption of the X-ray emission. Taking these 
considerations into account, we show that the situation is more
complicated than assuming a single power-law for the optical/X-ray spectrum. 
We have also done the same exercise for the PWN. 

The outline of the paper is as follows: in Sect. 2 we describe the optical
spectroscopic and photometric observations of \psr\ and its PWN, as well
as a reinvestigation of the X-ray data of Kaaret et al. (2001). In
Sect. 3 we discuss these results and put them in a multiwavelength context.
We also discuss results we find for the proper motion of the 
pulsar, and their possible interpretation.

%---------Observations and data analysis--------------------

\section{Observations, data analysis and results}
\subsection{VLT observations and data reduction}
%-----------------------table2------------------------------
%-----------------------table2------------------------------
\begin{table}[b]
\caption{Log of VLT observations of \psr~on 2002 January 9.}
\label{t:obs} 
\begin{tabular}{lcccc}
\hline
\hline
No.  &Time & Exposure &Airmass & Seeing$^a$  \\ 
     &    UT &   s      &            & arcsec \\
\hline
1    & 01:59:32.0 & 1320 & 1.43 & 1.25 \\
2    & 02:23:25.0 & 1320 & 1.42 & 1.30 \\
3    & 02:52:38.0 & 1320 & 1.41 & 1.18 \\
4    & 03:16:31.0 & 1320 & 1.41 & 1.12 \\
5    & 03:42:32.0 & 1320 & 1.41 & 1.24 \\
6    & 04:06:25.0 & 1320 & 1.43 & 1.09 \\
7    & 04:32:42.0 & 1320 & 1.45 & 0.88 \\
\hline
LTT 3864 & 08:51:03.0 & 20 & 1.03 & 0.67 \\
\hline
\end{tabular} \\
\begin{tabular}{lll}
$^a$ \  Full width at half maximum of the stellar profile.&& \\
\end{tabular}
\end{table}

\begin{figure}[tbh]
\begin{center}
\includegraphics[width=88mm, clip]{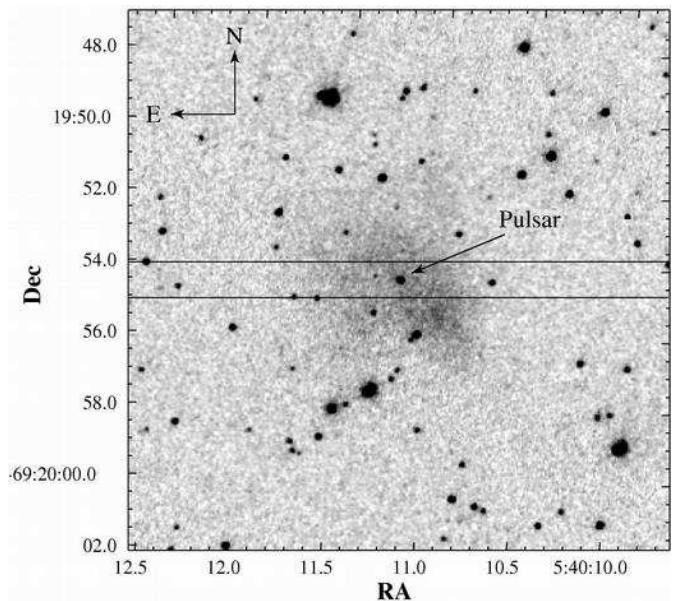}
\end{center}
\caption{A 15\arcsec$\times$15\arcsec\ image of the field
around \psr~obtained in the F547M band with the HST/WFPC2 (Morse 2003).
The pulsar was exposed on the PC chip and its position is marked by an
arrow. The slit position of the VLT observations is marked by thin
parallel lines. The slit width is 1\arcsec. The diffuse emission surrounding
the pulsar is the pulsar wind nebula (PWN). Note its elongation 
in the NE-SW direction.
} \label{f:F547im}
\end{figure}

%______________________________________________
Spectroscopic observations of \psr~were performed on 2002 January 9 with
the {\it FOcal Reducer/low dispersion Spectrograph} (FORS1) on the 
8.2m UT3 (MELIPAL) of the ESO/VLT, using a slit width of~1\arcsec~and the grism 
GRIS\_600B\footnotemark \footnotetext{\href{http://www.eso.org/instruments/fors1/grisms.html}
{http://www.eso.org/instruments/fors1/grisms.html}}. 
This grism has a dispersion of 50~\AA$/$mm, or 1.18~\AA$/$pixel, 
and a wavelength range of 3605--6060~\AA.  
The optical path also includes a Linear Atmospheric Dispersion Corrector that 
compensates for the effects of atmospheric dispersion (Avila et al. 1997). 
The pixel scale of the detector is 0\farcs2 per pixel. 
We obtained 7 exposures of 1320 s 
each (see Table~\ref{t:obs}), in total 154 minutes of exposure time.
The position angle, PA=88\degr, was the same in all these exposures. 
The slit crosses the pulsar and its PWN as shown in Fig.~\ref{f:F547im}. 
The mean seeing was $\sim$ 1\farcs15. 
%-----------------------figure1------------------------------
%------------------------------------------------------------

The spectroscopic images were bias-subtracted and flat-fielded using standard 
procedures within the NOAO {\sf IRAF Longslit} package. 
We used the averaged sigma clipping algorithm {\sf avsigclip} with the 
{\sf scale} parameter set equal to {\sf none} to combine the images.
Wavelength calibration of the combined images was done using arc frames 
obtained with a helium-argon lamp.
The spectra of the objects were then extracted from the 2D image 
%-----------------------figure2------------------------------
\begin{figure}[t]
\includegraphics[width=70mm, clip]{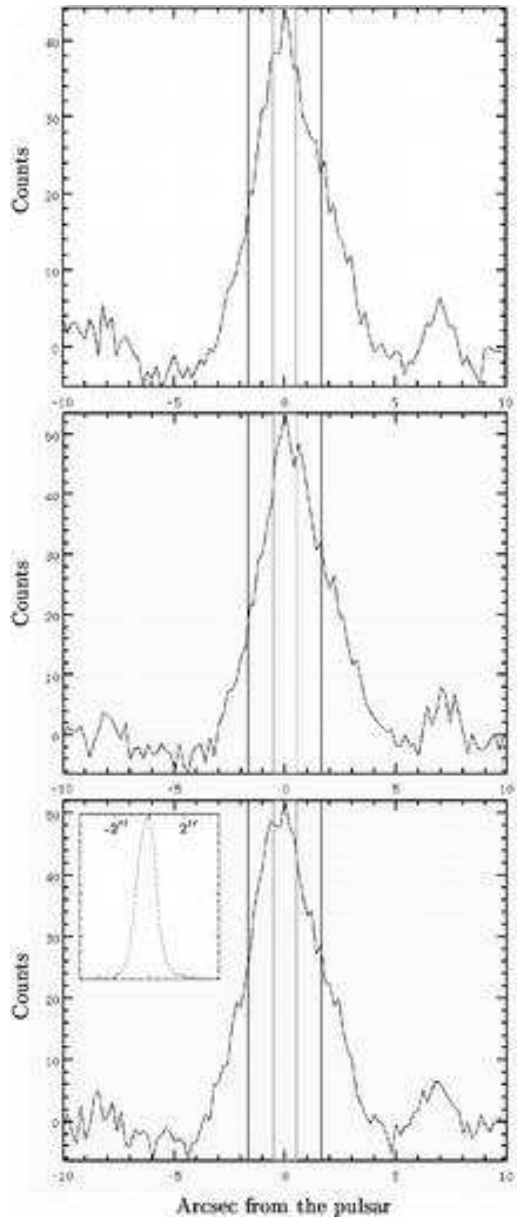}
\caption{Spatial profiles of \psr~and its PWN along the slit 
in Fig.~\ref{f:F547im} (in counts, VLT data)  
at the continuum wavelengths %$\lambda= 
4600.0~\AA, 5248.7~\AA\ and 5450.0~\AA, 
from top to bottom, respectively. In each panel the pulsar is assumed to sit 
at the main peak of the profile. Thin lines mark the six-pixel wide 
extraction window chosen for the spectral analysis of the pulsar (where 
1 pixel corresponds to 0\farcs2).
The regions outside the thin lines, but within the solid lines have also 
a width of six pixels, and mark the regions used for the nebular subtraction
in the spectral analysis. The spatial profile of a background star is shown in 
the inset in the bottom panel for comparison.
} 
\label{f:nebProfima}
\end{figure}
%----------------------------------------------
% ------------------ figure 3 ----------------
\begin{figure}[t]
\includegraphics[width=70mm, clip]{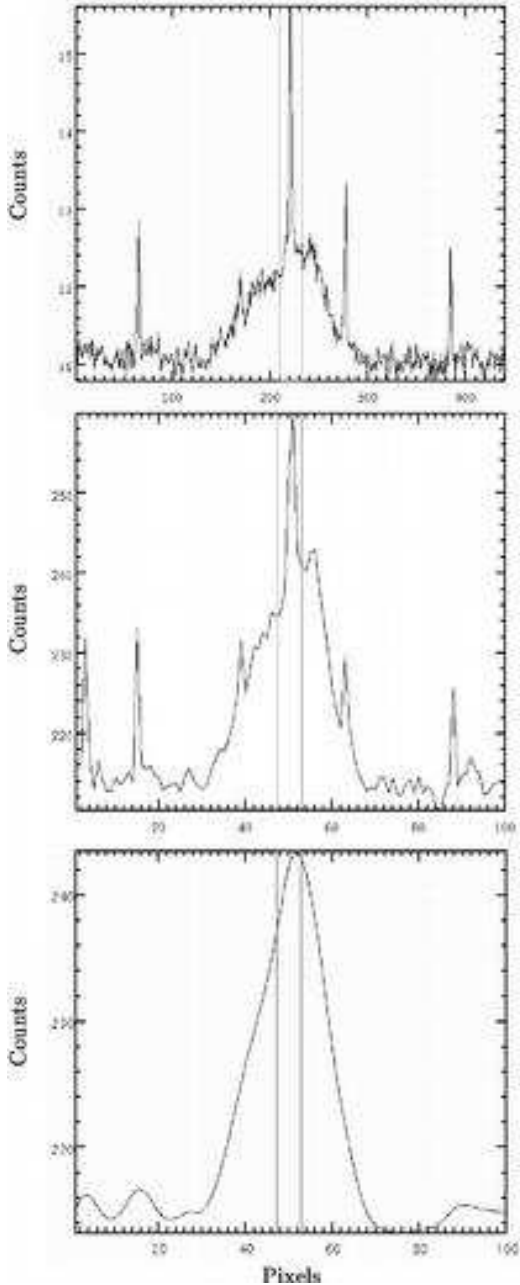}
\caption{Spatial profiles as in Fig.~\ref{f:nebProfima}, but using 
the HST/WFPC2/F547M image shown in  Fig.~\ref{f:F547im}. 
The pivot wavelength is 5483~\AA.      
We show profiles obtained along the VLT slit 
with the spatial resolution of 0\farcs046 provided by the PC chip 
pixel-size {\sl (top)}, rebinned to the VLT CCD-pixel
scale of  0\farcs2 {\sl (middle)}, 
and smoothed using a Gaussian with the FWHM of 1\farcs2,  
comparable to  the seeing value of the VLT observations, 
and then rebinned to the VLT CCD-pixel scale  
{\sl (bottom)}. The VLT six pixel extraction window is marked by vertical 
lines.
} 
\label{f:nebProfimb}
\end{figure}
%--------------------------------- 
 using the {\sf apall} and {\sf background} tasks.
Flux calibration of the spectra was accomplished by comparison with the 
spectrophotometric standard star LTT 3864 (\cite{ham94})
observed on the same night (see Table~2). 
Atmospheric extinction corrections were performed using 
a spectroscopic extinction table 
provided by ESO\footnote{http://www.eso.org/observing/dfo/quality/UVES/files/ \\ 
/atmoexan.tfits}. A significant flux (response) degradation  
was found  at wavelengths below 3860~\AA\, where the standard star
spectrum displays many deep spectral features, 
as well as in the very red end at $\lambda\ga$~6030~\AA.
We therefore excluded these wavelength regions from our analysis.   
%%%%%%%%%%%%%%%%%%%%%%%%%%%%%%%%%%%%%%%%%%%%
\subsection{Spectroscopy of the pulsar}
%%%%%%%%%%%%%%%%%%%%%%%%%%%%%%%%%%%%%%%%%%%%%%%
As seen from Fig.~\ref{f:F547im} the pulsar is in the center of a bright
compact nebula, which contaminates the flux from the pulsar. 
This contamination is particularly strong in nebular lines from 
the supernova remnant, but is also significant at continuum wavelengths  
where the spectrum of the extended object is expected to be 
dominated by synchrotron emission from the PWN.   

In Fig.~\ref{f:nebProfima}   
we have plotted spatial profiles for the emission from 
the pulsar+nebula along the slit at several 
wavelengths where the contribution of nebular lines is negligible.  
Despite some minor variation of the shape of the profile
with wavelength, the profile has a strong peak around the pulsar position.  
However, with a seeing of $\sim1\arcsec$ the 
pulsar and nebular emissions are strongly blended.
This is partially confirmed by a comparison of the profile 
with the PSF of a background star shown in the inset of
the bottom panel of Fig.~\ref{f:nebProfima}.
To illustrate this further we have compared the VLT profiles with the profiles obtained 
from the HST/WFPC2/F547M image, shown in Fig.~\ref{f:F547im}. 
From the F547M image we first extracted the data covered by the 
VLT slit, and then we averaged the emission across the slit for each position
along the slit.
The peak position of the flux along the slit varies with the wavelength of 
the spectroscopic image by $\lsim 1$ VLT pixel. 
To take this into account in our test we extended the slit width by one pixel. 
In the first test case we kept the high spatial resolution in the HST image 
(0\farcs046, Fig.~\ref{f:nebProfimb} top), 
while in the second case we averaged over a coarser pixel scale (0\farcs2) to
simulate the VLT pixel size (Fig.~\ref{f:nebProfimb} middle). 
In a third experiment we 
smoothed the initial image using a Gaussian with a FWHM of 1\farcs2 to model 
the VLT seeing conditions and we then rebinned it to the VLT CCD-pixel scale  
(Fig.~\ref{f:nebProfimb} bottom). 

The pivot wavelength of the F547M filter is $\sim 5483$~\AA, 
which is close to  $5450$~\AA, chosen for the profile 
shown at the bottom of Fig.~\ref{f:nebProfima}. 
As seen from Fig.~\ref{f:nebProfimb}, 
the pulsar is clearly resolved  
from the extended PWN at the PC chip spatial 
resolution as a narrow central peak 
on a broad ($\sim7$\arcsec\ in size)   
asymmetric pedestal formed by the PWN.     
It is still resolved at the VLT CCD-pixel scale of 0\farcs2, while it is not
resolved after the 1\farcs2 smoothing. i.e., close to the seeing conditions 
of the VLT observations. 
Although the pulsar should contribute significantly to the flux within 
the spatial strip of six VLT pixels 
centered on the main peak of the whole profile, it is obvious 
that the nebula will contaminate severely the spectral VLT observations.

\begin{figure}[t]
\begin{center}
\includegraphics[width=85mm, height=100mm, clip, angle=0]{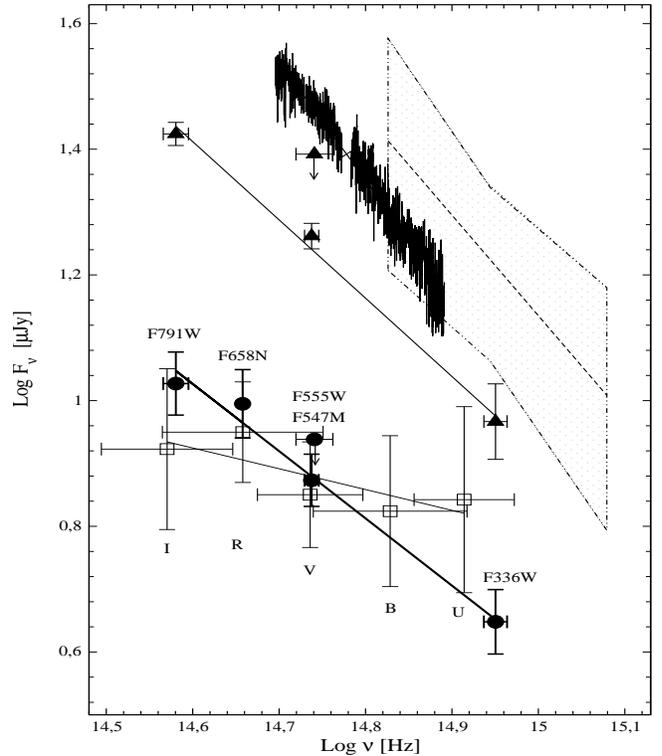}
\end{center}
\caption{Optical spectrum of \psr\ obtained with different telescopes and 
instruments. The uppermost spectrum is the VLT spectrum 
for the 6-pixel area discussed in Figs.~\ref{f:nebProfima} 
and \ref{f:nebProfimb}. 
The bright [O~III] nebular lines have been removed.  
The dashed line and associated hexagonal region 
show the power law fit and $1\sigma$-uncertainties 
of the UV spectrum obtained by Hill et al. (1997). 
Filled triangles show the HST photometry with a 10-pixel circular
aperture to compare with the above spectra.
Filled ellipses show our HST photometry results presented in 
Table~\ref{t:psr-phot}. 
Open rectangles are the photometric UBVRI data by Middleditch et al.~(1987).
All data are dereddened using $E(B-V)=0.20$. 
Solid lines show power law fits to the photometric data sets. 
Parameters of the fits are presented in Table~\ref{t:psr-fits}.  
} \label{f:PSRopt}
\end{figure}

With this in mind,
we extracted a 1D spectrum averaged over 6 pixels, equal to $1\farcs2$,   
along the slit centered at the pulsar position, as shown 
by thin vertical lines in Figs.~\ref{f:nebProfima} and \ref{f:nebProfimb}. 
To subtract the nebular contribution we extracted 1D spectra averaged over 
six adjacent pixels to the east and an equal number to the west of
the central strip, as indicated by thick lines in Fig.~\ref{f:nebProfima}.     
An averaged spectrum was constructed from the two adjacent spectra and
subtracted from the spectrum for the central region.
The resulting spectrum contains no significant emission from nebular lines, 
except for [O~III]~$\wll$4959,~5007,
which has a high spatial variability within the nebula.           
The resulting spectrum, with the [O~III]-feature  removed,
is presented in Fig.~\ref{f:PSRopt}. As can be seen from this 
figure, our spectroscopic data agree well with the result 
by Hill et al. (1997) at the lower boundary of their 1$\sigma$ 
uncertainty range, but give about 2$-$4 times  
higher flux than the photometric data of Middleditch et al. (1987).
This can partly be understood from our method of correcting for the
PWN emission. We averaged this by using the emission $1\farcs2 - 2\farcs4$ away
from the center of the spatial profile. One problem is 
that Fig.~\ref{f:nebProfima} 
shows that the center of the profile in the VLT data may not coincide exactly 
with the pulsar position. A second, and more serious problem, is that  
the seeing spreads out much of the weak pulsar emission from the central region 
whereas seeing makes the PWN emission peak toward the center regardless of
whether there is a pulsar or not. This will most likely lead to erroneous 
background subtraction so that the VLT spectrum shown in Fig.~\ref{f:PSRopt} is 
contaminated with significant PWN emission. 
The agreement with Hill et al. (1997)
indicates that also their analysis overestimated the pulsar emission.

%%%%%%%%%%%%%%%%%%%%%%%%%%%%%%%%%%%%%%%%%%%%%%%%%%%
\subsection{HST observations}
%%%%%%%%%%%%%%%%%%%%%%%%%%%%%%%%%%%%%%%%%%%%%
The pulsar field has been imaged  
with the HST/WFPC2 several times in various bands mainly   
to study SNR~0540-69.3. For additional analysis of the optical 
emission from \psr~and its PWN we retrieved some of these images 
from the HST archive.

The data using the wide and medium band filters 
F336W\footnotemark \footnotetext{http://www.stsci.edu/instruments/wfpc2/
/Wfpc2\_hand\_current/ch3\_opticalfilters2.html\#474439},
F547M, and F791W, obtained on 1999 October 17 with 600~s, 800~s, and 400~s 
total exposure times, respectively (Morse 2003), are 
particularly useful for the continuum emission analysis 
since these filters do not cover any bright emission lines from the LMC 
or the supernova remnant.
We also retrieved data sets for the narrow band  F658N and  wide band F555W 
filters, both of which were obtained on 1995 October 19 with 4000~s, 
and 600~s exposures, respectively  (Caraveo et al. 2000).

The pulsar and its PWN are clearly detected on the PC chip in all these 
images. This is illustrated in Figs.~\ref{f:F547im} and~\ref{f:nebProfimb} 
which present the data for the F547M band. 
The F658N filter includes the 6576$-$6604~\AA~range, 
which means that contamination from high-velocity H$\alpha$ emission 
from the supernova remnant at central wavelength $\sim 6578$~\AA\ 
(Serafimovich et al. 2004) can enter into the filter passband, as well as
[N~II] emission from the LMC. This can also be seen in Fig. 3b of
Caraveo et al. (2000) where it is shown that a filament passes across the 
pulsar along the NW direction. Although the pulsar stands out rather
clearly on the image, the uneven background introduces some uncertainty
to the estimated pulsar flux. The background contamination is more severe in
the F555W band as it captures the bright [O~III]$\wll$4959,~5007~\AA\ lines 
which are much more difficult to spatially disentangle from the pulsar. 
The uneven background is most clearly seen in the image obtained with the 
F502N narrow band filter centered at these lines and overlapping with the 
F555W band. Therefore, the pulsar and PWN continuum flux measurements 
in the F555W band can only be considered as upper limits.
%-----------------------figure4------------------------------

%%%%%%%%%%%%%%%%%%%%%%%%%%%%%%%%%%%%%%%%%%%%%%%%%%%
\subsection{Photometry of the pulsar}
%%%%%%%%%%%%%%%%%%%%%%%%%%%%%%%%%%%%%%%%%%%%%
For the photometry  of the pulsar in the 
F336W, F547M, F555W, F658N, and F791W bands 
we used the {\sf Daophot and Photcal} {\sf IRAF} packages. 
To measure the pulsar flux  we used circular 
apertures whose radii  were  determined  from maximizing   
the signal-to-noise ratio S/N of the detected pulsar,  
and hence minimizing the magnitude error 
$\Delta m$  in each filter. For a point-like source, $\Delta m$ and S/N
are defined in the standard way as
\begin{equation}
\Delta m 
=\frac{2.5}{\ln 10} \left( \frac{S}{N} \right)^{-1}\label{eq1}, 
\end{equation} 
\begin{equation}
\frac{S}{N}=\left[\frac{N_c}{G} + 
A\sigma_N^2(1+\frac{A}{A_{sky}})\right]^{-1/2}N_c,\label{eq2}
\end{equation}
where $N_c$ is the flux in counts for a given aperture, 
$\sigma_N^2$ is the standard deviation in counts per pixel, 
$A$ is the number of pixels in the aperture, 
$G$ is  the gain, and $A_{sky}$ is the number of pixels 
in the annulus  used for background  measurements.  
The optimal aperture radii were found to be 
2.5, 3.0, 4.0, 2.5, and 3.5 pixels, providing S/N of 8, 10, 25, 
7.5, and 8 in the F336W, F547M, F555W,  F658N, 
and F791W bands, respectively. 
Aperture  corrections for the magnitudes    
were derived with the  {\sf mkapfile} package 
using   several relatively bright nearby 
field stars in the PC images. 

Pipeline-provided zeropoints 
PHOTFLAM (flux densities in wavelength units) and  
pivot wavelengths PHOTPLAM (in \AA)   taken from 
the fits header of each image were used for the flux calibration
(see handbook for WFPC2\footnotemark \footnotetext
{www.stsci.edu/instruments/wfpc2/
/Wfpc2\_hand\_current/ch8\_calibration9.html\#464103}).
Fluxes in units of {\rm $\mu$Jy} 
at the pivot frequencies 
$\nu=10^8$c/PHOTPLAM (in {\sf Hz})   
were derived from the aperture corrected source counts using the expression  
\begin{equation}
F_{\nu} = 10^{21}\frac{N_c}{c~t_{exp}} PHOTFLAM \times PHOTPLAM^2 
\label{eq3}    
\end{equation} 
where $c$ is the speed of light in {\sf cgs} units, and 
$t_{exp}$ is the exposure time in seconds.    
We also calculated dereddened fluxes by multiplying 
$F_{\nu}$ with 
the extinction correction factors 
$k_{ext} = 10^{0.4 k_i}$, 
where $k_i$ is the interstellar extinction 
magnitude for the  $``i$-$th''$  passband  
determined using the extinction color excess 
toward SNR~0540-69.3. In order to compare with Middleditch et al. (1987)
we have used the same values for $E(B-V)$ and $R$, i.e., 0.20 and 3.1,
respectively, for both data sets, but  
a more recent extinction curve (Cardelli et al. 1989).
We will discuss the extinction in greater detail in Sect. 2.8.

The results of the pulsar photometry are   
presented in Table 3 and  shown in Fig.~\ref{f:PSRopt}.    
The measured fluxes are about a factor of 2$-$4 
lower than the spectroscopic results in Sect. 2.2, 
except for the F555W filter which 
includes contamination from [O~III].  The fluxes in this filter 
are presented as upper limits (see above). 
On the other hand, the HST photometry 
is compatible with the results obtained by Middleditch et al. (1987), 
but the uncertainties  are several times smaller. 
%%%%%%%%%%%%%%%%%%%%%%%%%%%%%%%% Table3, photometry %%%%%%%%%%%%%%%%%%%%%%%%%  
\begin{table}[t]
\caption{Broad-band fluxes$^a$ from \psr.}
\label{t:psr-phot} 
\begin{tabular}{llcc}
\hline
\hline
Band    &  Pivot  & Measured    & Dereddened            \\ 
        &frequency  & flux   &    flux           \\ 
        &  log $\nu$~[Hz]    & log $F_{\nu}$~[{\rm $\mu$Jy}]    & log $F_{\nu}$~[{\rm $\mu$Jy}]       \\ 
%       &   [Hz]    & [{\rm $\mu$Jy}]    &  [{\rm $\mu$Jy}]    \\          
\hline 
F336W    & 14.950(13)        &   0.243(51)   & 0.648(51) \\ 
F547M    & 14.737(8)         &   0.625(42)   & 0.873(42)\\ 
F555W    & 14.741(21)        &$\le$ 0.69     & $\le$ 0.94  \\  
F658N    & 14.658(1)         & 0.793(54)     & 0.995(54)    \\
F791W    & 14.581(15)        & 0.874(50)     & 1.027(50)   \\
\hline
\end{tabular} \\
\begin{tabular}{ll}
$^a$~Hereafter numbers in the parentheses are uncertainties 
& \\
 \ of the respective values. For example, 14.950(13) means & \\
 \ $14.950\pm0.013$.& \\ 
%1$\sigma$ flux uncertainties related to last digital    .
\end{tabular}
\end{table} 
%%%%%%%%%%%%%%%%%%%%%%%%%%%%%%%% Table 4, 10-pix photometry %%%%%%%%%%%%%%%%%%%%%%%%%  
\begin{table}[t]
\caption{Broad-band fluxes from a 10 pixel aperture around \psr.}
\label{t:10pix}
\begin{tabular}{llcc}
\hline
\hline
Band    &  Pivot  & Measured    & Dereddened            \\ 
        &frequence  & flux   &    flux           \\ 
        &  log $\nu$~[Hz]    & log $F_{\nu}$~[{\rm $\mu$Jy}]    & log $F_{\nu}$~[{\rm $\mu$Jy}]       \\ 
%       &   [Hz]    & [{\rm $\mu$Jy}]    &  [{\rm $\mu$Jy}]    \\          
\hline 
F336W    & 14.950(13)         &   0.562(60)   &   0.967(60)          \\ 
F547M    & 14.737(8)         &   1.013(20)   &    1.262(20)    \\ 
F555W    & 14.741(21)        &$\le$ 1.141      & $\le$ 1.392    \\  
%F658N    & 14.658(1)         & 0.793(54)     &0.995(54)       \\
%%% why is this not included?? jesper
F791W    & 14.581(15)        & 1.271(18)      & 1.424(18) \\
\hline
\end{tabular} 
\end{table} 
%%%%%%%%%%%%%%%%%%%%%%%%%%%%%%%%% 
%%%%%%%%%%%%%%%%%%%%%%%%%%%%%%%% Table 5, spectral fits of data in Fig.4 %%%%%%% 
\begin{table}[tbh]
\caption{Parameters of the power law spectral fits ($F_{\nu}=F_{\nu_0}(\nu/\nu_0)^{-\alpha_{\nu}}$) 
of the pulsar data shown in Fig.~\ref{f:PSRopt}.}
\label{t:psr-fits} 
\begin{tabular}{lcc}
\hline
\hline
Observations   &  $\alpha_{\nu}$    &  Log~$F_{\nu_0}$, \\
& & $\nu_0$=$5.47\times10^{14}$ Hz         \\ 
        &   &  [{\rm $\mu$Jy}]                \\             
\hline 
VLT spectrum        &  1.88$\pm$0.01 & 1.469(2)  \\ 
(this work)         &    & \\ 
HST/FOS spectrum$^a$&  1.6$\pm$0.4   & 1.228(115)\\ 
(Hill et al. 1997)  &    & \\  
HST photometry      &  1.07$^{+0.20}_{-0.19}$ & 0.879(25) \\  
(this work)         &    & \\ 
10 pixel measurements & 1.24$^{+0.19}_{-0.18}$ & 1.241(22)\\ 
(this work)         &    & \\  
Time-resolved photometry   & 0.26$^{+0.45}_{-0.43}$ & 0.880(54)\\ 
(Middleditch et al.~1987)  &    & \\  
\hline
\end{tabular}\\ 
$^a$~Log~$F_{\nu_0}$ from Hill et al. (1997).\\
\end{table} 
Considered separately, our spectral and photometric measurements of the pulsar 
flux are in good agreement with previous results. However, they  do not erase
the significant discrepancy between these results (see Fig.~\ref{f:PSRopt}), 
which is much 
larger than the uncertainty of our photometric measurements. The only 
plausible explanation to the discrepancy is that both the VLT and HST 
spectroscopy are strongly contaminated by the PWN, as already suggested in 
Sect. 2.2. As an additional test we  
measured the pulsar flux in the F336W, F547M, F555W, and F791W bands, using  
a circular aperture with a radius of 10 PC-pixels (corresponding to a 
total diameter of
0\farcs92) centered on the pulsar, without subtraction of the background. 
These conditions should approximately reproduce the parameters of the spectral 
measurements within a circular aperture of almost the same diameter (0\farcs86) 
made with the HST/FOS by Hill et al. (1997). The measured fluxes 
are presented in Table~\ref{t:10pix} and 
shown by triangles in Fig.~\ref{f:PSRopt}. They are much 
closer to the HST and VLT spectral fluxes.
The $\sim$20\%--70\% discrepancy between these HST/WFPC2  
and FOS fluxes must be considered as small compared with the large 
uncertainties of the FOS flux.
Using a WFPC2 image Hill et al.~(1997) estimated the PWN
contribution to be 30\% in their FOS aperture. Note that this has not been
accounted for in the spectrum in Fig.~\ref{f:PSRopt}. However, their estimate assumed a
uniform nebular background and was performed on an image with a PSF
affected by spherical abberation. 
We have redone the exercise with the
F574M image from 1999, and estimate that within a 10 pixel radius, the PWN
contributes at least 50\%.
This can also be seen by comparing our accurate pulsar photometry
with our 10 pixel test. The contamination of the HST/FOS spectrum would
be even larger if the centering was not perfect.
Although formal uncertainties of the VLT spectral flux appear to be smaller, 
the contamination from the nebula in the optical range 
is even stronger than in the UV, exceeding 250\% at the redmost wavelengths
of the VLT spectral range. This is natural, 
since the brightness of the PWN increases more steeply with the 
wavelength than does the pulsar brightness (see below).   

%-----------------------figure5------------------------------
\begin{figure}[hbt]
\begin{center}
\includegraphics[width=81mm, clip, angle=0]{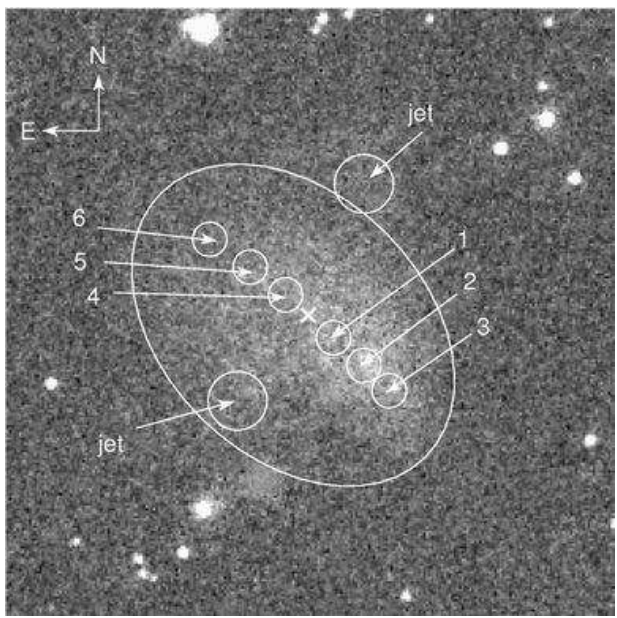}
\end{center}
\caption{10\arcsec$\times$10\arcsec\ region of the field around \psr~as
observed in the F547M band with the HST/WFPC2 (Morse 2003). 
Background stars and the pulsar projected on the PWN  have been subtracted 
off (cf. Fig.~\ref{f:F547im}). 
We have marked six selected areas in the plane of the assumed torus as well
as two areas in the jets. The ellipse shows the aperture used to measure the
flux from the whole PWN. The pulsar position is marked by a white  cross.  
} \label{f:PWNstady}
\end{figure}
%------------------------------------------------------------
To summarize this part, we conclude that the current spectral 
data on the pulsar emission cannot be considered as reliable. 
Hopefully, narrow-slit spectroscopy with the HST/STIS
could help to solve the contamination 
problem. Time resolved photometry with a high signal-to-noise is also a 
powerful tool to obtain the pulsed emission in a contaminated area, as can 
be seen even for the Crab pulsar (Sollerman 2003).
On the other hand, our  broadband HST spectrum 
of the pulsar, where the background from the nebula has 
been accurately subtracted off,   
can be considered as a fair estimate of the pulsar spectral energy 
distribution. As seen from Fig.~\ref{f:PSRopt}, it is significantly steeper
than found by Middleditch et al. (1987). If we define the spectrum 
as $F_{\nu}=F_{\nu_0}(\nu/\nu_0)^{-\alpha_{\nu}}$, then our power-law index
is $\alpha_{\nu} =1.07_{-0.19}^{+0.20}$ while that of Middleditch et al. 
is $\alpha_{\nu} =0.33\pm0.45$ (see Table~\ref{t:psr-fits}) using updated 
dereddening corrections. (The method how to calculate these power-law
indices is described in Sect. 2.5.) 
The flatter spectrum of Middleditch et al. is partially due to a spectral 
jump upwards for the U band, whereas our flux toward the overlapping F336W 
band goes down smoothly, as would be expected 
from the extrapolation of the longer wavelength data. This upturn in 
Middleditch et al. (1987) could be due to a systematic error in their U band 
flux. We note that also the Crab-pulsar broadband spectrum by Middleditch et 
al. (1987) has a significant excess in the U band, which has not been 
confirmed by more recent spectral observations extending even further into 
the UV (Sollerman et al.~2000). 
However, possible variability of the pulsar 
emission in the UV range cannot be excluded as another cause of the 
difference between our results for \psr\ and the result by Middleditch 
et al. (1987). We also note that while we are studying the time-integrated 
flux, Middleditch et al. studied the time-resolved flux, 
albeit with a poor signal-to-noise and spatial resolution. 

%%%%%%%%%%%%%%%%%%%%%%%%%%%%%%%%%%%%%%%%%%%
\subsection{Photometry of the PWN}
%%%%%%%%%%%%%%%%%%%%%%%%%%%%%%%%%%%%%%%%%%%%%%
Using the HST images discussed in Sect. 2.3 we have also made 
aperture  photometry of the continuum emission from the whole PWN, as
well as different parts of it selected on the basis of the morphology 
of the nebula. For this analysis we have discarded the F555W and F658N 
filters due to the risk of contamination by the [O~III] and H$\alpha$ emission 
from the SNR. To make an accurate analysis we first subtracted off 
all stars in the 
%-----------------------figure6------------------------------
\begin{figure}[htb]
\begin{center}
\includegraphics[width=65mm, clip, angle=270]{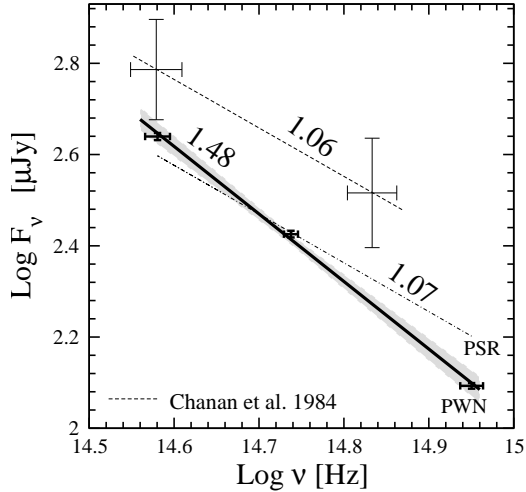}
\end{center}
\caption{Broad-band fluxes from the full PWN obtained with the HST 
(thick errorbars) using the elliptical aperture shown in Fig.~\ref{f:PWNstady}. 
The thick line and filled area provide the best power-law fit to the %HST 
data and its 1$\sigma$ uncertainties, respectively (see  Table~\ref{t:PWN-phot}). 
Previous data by Chanan et al.~(1984) in the I and B bands 
(thin errorbars and dashed line), as well as the slope of the pulsar spectrum
(cf. Fig. 4), normalized arbitrary (dot-dashed line), 
are shown for comparison. Numbers near the lines denote spectral indices.
Note that Chanan et al. (1984) did not correct for contaminating flux from
background stars and nebular line emission.
} 
\label{f:PWNellip}
\end{figure}
%------------------------------------------------------------
immediate vicinity of the nebula as well as stars which overlap with
the PWN. We also subtracted off the pulsar. For the subtraction we used 
the {\sf allstar} task within the {\sf Daophot} package (\cite{st87}) 
and for a comparison also the {\sf credit} task within the {\sf Crutil} 
package. There was no significant difference in 
the subtraction  results  using the two methods. 
The photometry was performed with the {\sf Isophote} package (\cite{Jer87}).  

The PWN of \psr\ has a remarkable structure that can be seen clearly in 
Figs.~\ref{f:F547im} and \ref{f:PWNstady}. To sample different regions,  
we have put circle apertures at eight 
different positions, one in each jet (with an aperture radius of 12 pixels, 
i.e. 0\farcs55), and three (with an aperture radius of 7 pixels, i.e. 0\farcs32) on each 
side of the pulsar in the plane of the presumed torus. We also constructed 
an elliptical aperture with 
a 74 pixel (i.e.~3\farcs4) semi-major axis, ellipticity 0.3, and positional 
angle 45 degrees,   
that encapsulates the entire PWN (cf. Fig.~5), except for the weak northern jet. 
The apertures are marked and identified in Fig.~\ref{f:PWNstady}. 
All regions  show emission 
in all filters, except for the northern jet (``North Jet'') and ``Area 6'' 
which are not detected in the F336W band. For these two regions we 
provide 3$\sigma$ upper limits, based on the standard 
sky deviations per pixel within the respective areas.  
The results are presented in Table~\ref{t:PWN-phot}.

%%%%%%%%%%%%%%%%%%%%%%%%%%%%%%%% Table 6, PWN  photometry %%%%%%%%%%%%%%%%%%%%%%%%%  
\begin{table*}[hbt]
\caption{Broad-band  fluxes 
and power-law fit ($F_{\nu}=F_{\nu_0}(\nu/\nu_0)^{-\alpha_{\nu}}$) 
parameters of the emission from different regions of the PWN marked in Fig.~\ref{f:PWNstady}.  
Their offsets from the pulsar and areas are given in the 2nd column. 
Upper and lower 
values for each entry of the fluxes are the measured and dereddened fluxes 
with $E(B-V)=0.20$ ($A_V=0.62$), respectively. }
\label{t:PWN-phot}
\begin{tabular}{lllcccccccc}
\hline
\hline
\multicolumn{3}{c}{Source}&  &\multicolumn{3}{c}{Band} & &\multicolumn{2}{c}{Power-law fit} \wideru \\
\cline{1-3} \cline{5-7} \cline{9-10} 
%\hline
      &   &   &  &    &    &   & & &         \\ 
Region   & Offset & Area &  & F336W    & F547M  &  F791W & &$\alpha_{\nu}$&${\rm log}~F_{\nu_0}$, $\nu_0=5.47\times10^{14}$ Hz   \\
         &        &arcsec$^2$ &  & \multicolumn{3}{c}{ Log flux from the area~~~{\rm $\mu$Jy} }      & & &  {\rm $\mu$Jy}       \\  
\cline{1-3} \cline{5-7} \cline{9-10}         
%\hline 
      &   &     &  &  &    &   & & &         \\ 
Full PWN   &  0\farcs000 N  &10.92  &    &   1.688(6)   &   2.177(7) & 2.482(8)  & & --  & --         \\
             &  0\farcs000 W  &       &    &   2.093(6)   &   2.426(7) & 2.640(8)  & & 1.48$^{+(9)}_{-(8)}$  & 2.414(12)   \\
      &   &     &  &  &    &   & & &         \\ 
Area 1           &  0\farcs455 S   &  0.33  &  & 0.350(116)   &0.647(94)    &0.991(48)   & & --  &  --        \\ 
             & 0\farcs409  W   &        &  &0.755(116)   & 0.896(94)   &1.144(48)   & &1.09$^{+(33)}_{-(30)}$ &  0.949(50)   \\  
      &   &     &  &  &    &   & & &         \\        
Area 2           &  1\farcs001 S   &  0.33  &  & 0.224(146)   &0.766(35)    &1.054(30)   & & --  &  --        \\ 
             & 0\farcs910  W   &        &  &0.629(146)   & 1.015(35)   &1.207(48)   & &1.58$^{+(43)}_{-(42)}$ & 0.980(45)    \\
      &   &     &  &  &    &   & & &         \\        
Area 3           &  1\farcs456 S   &  0.33  &  & 0.176(147)   &0.628(60)    &0.892(45)   & & --  &  --        \\ 
             & 1\farcs365  W   &        &  &0.581(147)   &0.877(60)   &1.045(45)   & &1.28$^{+(43)}_{-(41)}$ &  0.856(49)   \\
      &   &     &  &  &    &   & & &         \\        
Area 4           &  0\farcs409 N   &  0.33  &  & 0.114(236)   &0.582(128)    &0.868(102)   & & --  &  --        \\ 
             & 0\farcs364 E   &        &  &0.519(236)   &0.831(128)   &1.021(102)   & &1.49$^{+(67)}_{-(59)}$ &  0.806(58)   \\
      &   &     &  &  &    &   & & &         \\        
Area 5           &  1\farcs046 N   &  0.33  &  & 0.252(140)   &0.485(106)    &0.791(64)   & & --  &  --        \\ 
             & 0\farcs865 E   &        &  &0.657(140)   &0.734(128)   &0.944(64)   & &0.92$^{+(37)}_{-(33)}$ &  0.786(58)    \\
       &   &     &  &  &    &   & & &         \\        
Area 6          &  1\farcs775 N    &  0.33  &  & $\le$ 0.032$^a$ &0.239(98)   &0.377(102)   & & --  &  --        \\ 
             & 1\farcs365 E   &        &  & $\le$ 0.437     &0.488(98)   &0.530(102)   & &0.27$^{+(91)}_{-(90)}$ &  0.488(98)   \\
      &   &     &  &  &    &   & & &         \\   
North Jet    & 1\farcs001 N    &  0.96  &  & $\le$ 0.265$^a$ &0.642(50)   &0.875(68)   & & --  &  --        \\ 
             & 2\farcs366 W    &        &  & $\le$ 0.670     &0.891(50)   &1.028(68)   & &0.87$^{+(56)}_{-(54)}$ &  0.891(51)   \\
       &   &     &  &  &    &   & & &         \\   
South Jet    & 1\farcs274 S    &  0.96  &  & 0.483(128)     &0.796(43)   &1.105(33)   & & --  &  --        \\ 
             & 1\farcs547 W    &        &  & 0.888(128)     &1.045(43)   &1.258(33)   & &1.03$^{+(35)}_{-(30)}$ &  1.077(38) \\  
\hline
\end{tabular}\\ 
$^a$~3$\sigma$ upper limit.
\end{table*} 
%%%%%%%%%%%%%%%%%%%%%%%%%%%%%%%%%%%%%%%%%%%%%%%%%%%%%%%%%%%%%%%%%%%%%%%%%%%   
%-----------------------figure7------------------------------
\begin{figure*}[hbt]
\begin{center}
\includegraphics[width=110mm, clip, angle=270]{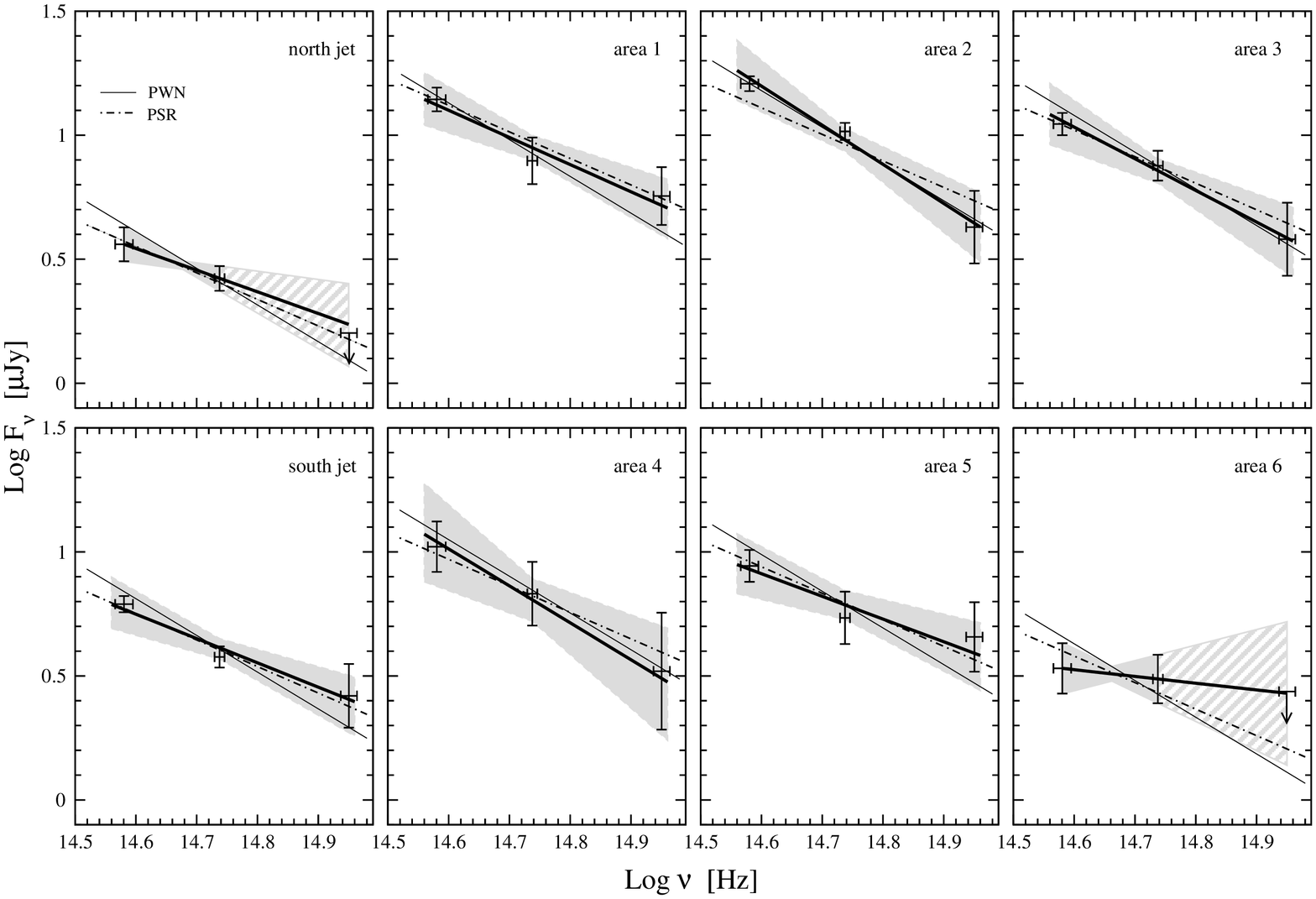}
\end{center}
\caption{Broad-band optical spectra for all the selected areas of the PWN  
shown in Fig.~\ref{f:PWNstady}. The filled hexagons show 1$\sigma$ uncertainty 
regions around the best-fit power laws indicated by thick lines. 
The  fluxes of the ``jets" are normalized to a 7-pixel aperture 
area for convenience in order to compare their brightnesses  
with the brightness of other parts of the PWN. 
The spectral slopes of the whole PWN (thin lines) and the pulsar 
(dot-dashed lines) are shown for comparison. The dot-dashed and thin lines
have been shifted arbitrarily in the vertical direction.
} 
\label{f:PWNstopt}
\end{figure*}
%------------------------------------------------------------
%------------------------------------------------------------     
    
As expected, the measured broadband spectra from the whole PWN and its 
different parts are well described by power-laws with negative spectral index,  
which confirm the nonthermal origin of the continuum nebular emission.
To derive the spectral indices we used the following method. 
For each data point we made Gaussian fits in log space to the flux and
the filter function. We then simulated 10,000 sets of data using a 
Monte Carlo code which uses a fast portable random 
generator\footnotemark \footnotetext{http://www.ntnu.no/$\sim$ joern/t3e-asm/vranmar.html}, 
and for each data set we made a linear fit to obtain a power-law.
The power laws were then ordered in increasing value of the power law index, 
and the median value was chosen to represent the best fit power law. In 
order not to be dependent on the seed value for the random number series, 
we ran the code 500 times with different seed values, and then took the 
average value for the power-law index to be the final estimate of the 
index. The filled hexagons in Figs. \ref{f:PWNellip} and \ref{f:PWNstopt} 
show 1$\sigma$ errors estimated from the constraint that 68\% of the 
constructed power laws must lie within a 1$\sigma$ area. 
The advantage of using a Monte Carlo code rather than simple weighted means to 
estimate power law indices is that we can allow and test for non-Gaussian 
distributions in log space. This is obviously not the case with a steep 
spectrum, non-Gaussian filter functions, as well as upper limits. 
Our tests, however, show that
the filters are narrow enough to get good fits from Gaussian fits to
the points with estimated fluxes. From our Monte Carlo code approach it is also 
easy to estimate the error of the derived power-law index. The same approach 
was also used to fit the pulsar spectrum in Sect.~2.4.    
The results are presented in 
Tables~\ref{t:psr-fits}, \ref{t:PWN-phot} and shown  
for the emission from the whole PWN in Fig.~\ref{f:PWNellip} 
and from its different regions in  Fig.~\ref{f:PWNstopt}.

According to Chanan et al. (1984) the fluxes (in $\mu$Jy)
from the whole PWN are ${\rm log}~F_B\approx 2.19$ 
and ${\rm log}F_I\approx 2.63$. Our values are 
much lower ($\sim$60\% and $\sim$40\%, respectively, see Fig.~\ref{f:PWNellip}).
The main reason is that we have subtracted off the pulsar 
and stars overlapping with the PWN, whereas these objects are not resolved 
from the PWN in the B and I images of Chanan et al. (1984) which were obtained 
at 1\farcs2-1\farcs4 seeing. Hence, their B and I fluxes are contaminated  
by non-PWN emission and this changes significantly the derived spectral slope 
of the PWN: Chanan et al. (1984) obtain  $\alpha_{\nu} = 1.06$ 
(after dereddening
with $E(B-V)=0.20$) whereas we get $\alpha_{\nu} = 1.48_{-0.08}^{+0.09}$ from 
the HST data. 

Compared with the pulsar the whole PWN is more than an order of magnitude 
brighter, and its spectrum is significantly softer 
(cf. Tables~\ref{t:psr-phot} and \ref{t:PWN-phot} and thick 
solid and thin dot-dashed lines in Fig.~\ref{f:PWNellip}). This shows that 
the NS spindown power is transformed to optical emission more efficiently   
in the PWN than in the pulsar magnetosphere (see below).       
   
A more detailed study shows that the spectrum may vary
over the PWN. This is seen from 
Fig.~\ref{f:PWNstopt} where we have plotted the results of the HST photometry 
of different parts of the PWN and the respective power-law spectral fits
(thick lines) with their $1\sigma$ uncertainties (uniformly filled regions). 
Stripe-filled regions show extensions of the fits in cases when 
only upper limits in one of the three bands were obtained.   
The spectral hardness of some parts of the nebula is comparable or even higher 
to that of the pulsar, as it is  for ``Areas 1, 5, 6", and  both  ``jets". 
The spectra of the N-E part of the torus-like structure appear to become  
harder toward the outer boundary of the nebula.    
On the other hand, ``Area 2", which is the brightest 
among the three selected areas S-W of the pulsar, has a steeper spectrum 
than its closest neighbors.
  
Another feature of the spatial flux and spectral variations 
of the PWN is demonstrated by Figs.~8 and 9 where we have 
plotted the distributions of the frequency-integrated optical fluxes $F$,  
derived from the above spectral fits, and the spectral  
indices $\alpha_{\nu}$ versus the angular distance from the pulsar 
along the major axis of the torus-like structure of the PWN. 
There is a significant decrease of the surface brightness going from 
the brightest area, area 2, toward the N-E edge of the torus (Fig.~8). 
The brightness difference exceeds the 6-sigma level of the uncertainty level 
of the dimmest area, area 6, and shows an asymmetry of the flux distribution 
with respect to the pulsar position which is similar to what is also seen in 
X-rays with Chandra/HRC\footnote{chandra.harvard.edu/photo/2004/snr0540}. 
In the Crab PWN, the brightness difference between the near and far sides of 
the torus, for a given viewing angle, is usually explained by Doppler 
boosting and relativistic aberration of the synchrotron radiation from 
relativistic particles flowing at subrelativistic velocities from the pulsar  
in the torus plane, assuming an axisymmetric distribution of the pulsar wind 
around the pulsar rotation axis (e.g., \cite{kl04}). 
However, the considerable asymmetry of the brightness distribution 
between the two sides (N-E versus S-W) of the torus of \psr, as seen in
projection, makes axial symmetry less obvious in a general picture. 
This is further strengthened by a similar asymmetry in the torus-plane, 
albeit less pronounced, seen in recent X-ray images of the Crab PWN 
(e.g., Mori et al 2004). The asymmetry can be produced either 
by breakdown of axial symmetry in the pulsar wind (e.g., due to plasma 
instabilities) or by inhomogeneity of the PWN environment, i.e.,
an asymmetry of the SN ejecta. The latter is indeed indicated by the 
asymmetric distribution of optical filaments projected on the PWN of \psr\ 
(\cite{Mo03}), as well as the general redshift of the gas emitting optical 
lines (Kirshner et al. 1989; Serafimovich et al. 2004).
%-----------------------figure8------------------------------
\begin{figure}[t]
\begin{center}
\includegraphics[width=49mm, clip, angle=270]{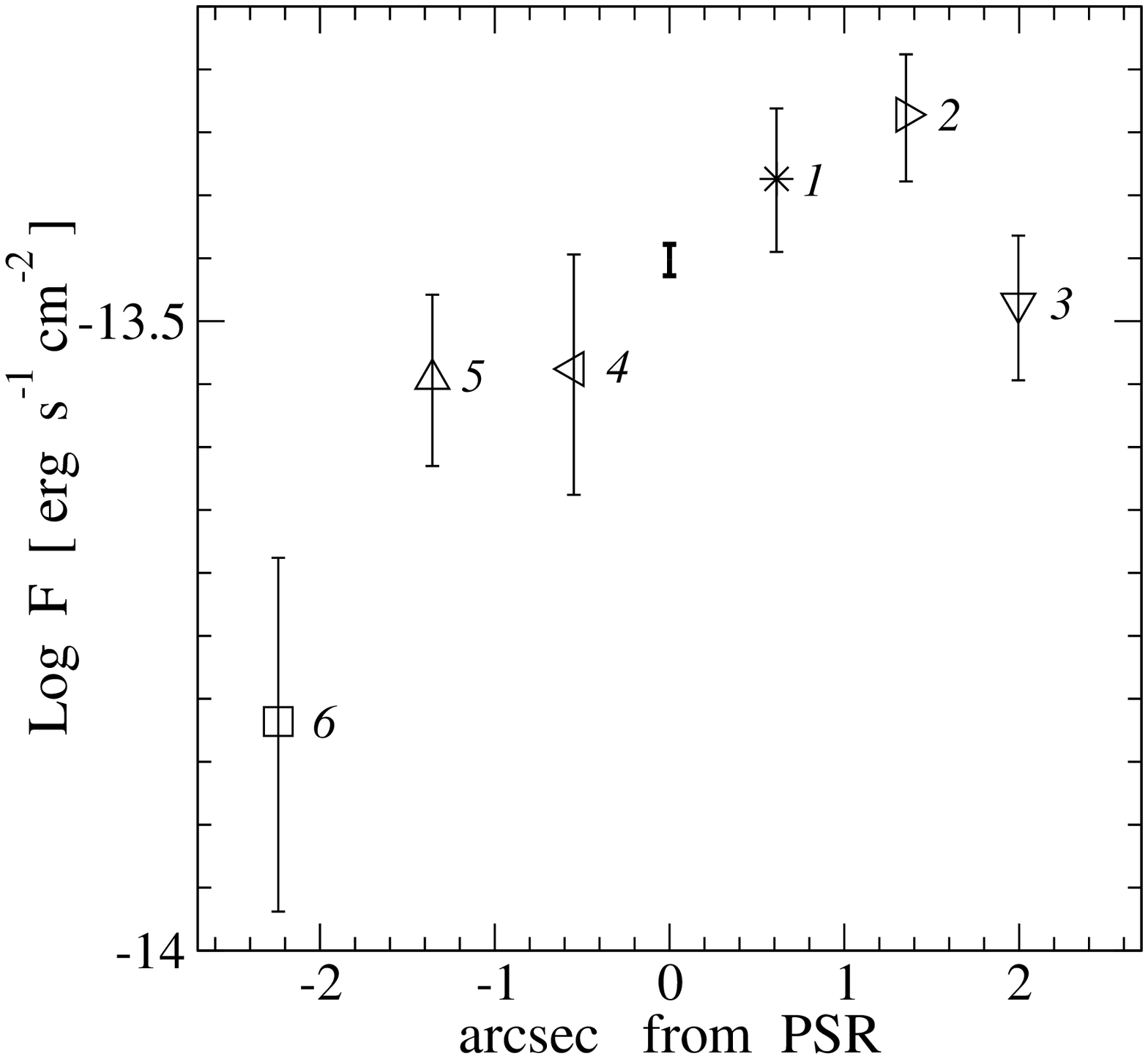}
\end{center}
\caption{Optical fluxes $F$ in the $(3.8-8.9)\times10^{14}$ Hz frequency
range from different parts of the PWN marked according to notations in 
Fig.~\ref{f:PWNstady}  
{\it versus} the distance from the pulsar along the major axis of the 
torus-like structure of the PWN. 
Negative and positive labels on the horizontal axis correspond to the N-E 
and S-W directions from the pulsar, respectively. The  bold errorbar at the 
center shows the flux from the whole PWN rescaled to the 7 pixel aperture 
area used for measurements in ``Areas 1 -- 6".    
} 
\label{f:PWN-fdist}
\end{figure}
%------------------------------------------------------------ 
%-----------------------figure9------------------------------
\begin{figure}[t]
\begin{center}
\includegraphics[width=49mm, clip, angle=270]{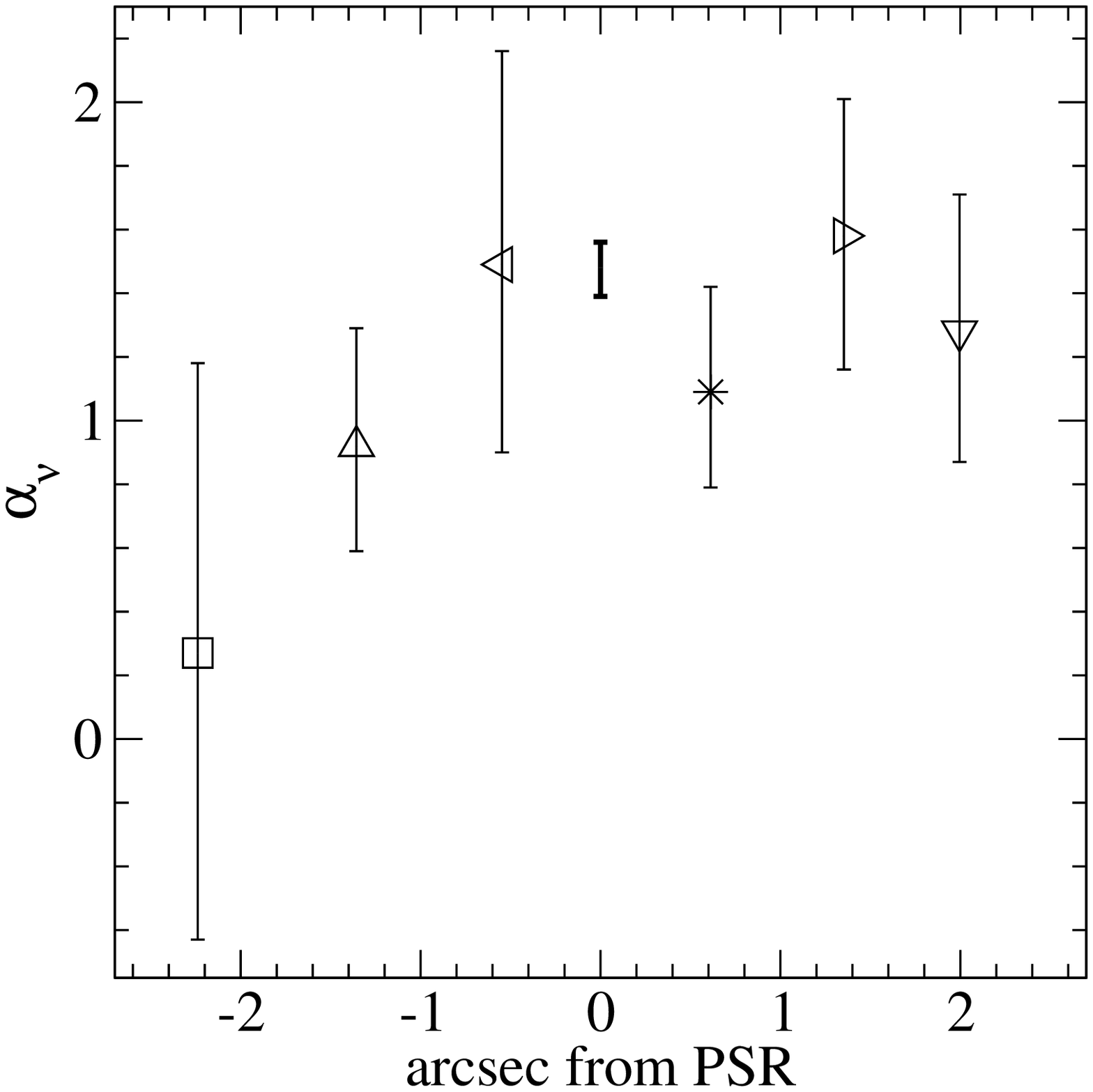}
\end{center}
\caption{Same as in Fig.~\ref{f:PWN-fdist} but for spectral indices. 
Symbols denote the same area numbers as in Fig.~\ref{f:PWN-fdist}.  
} 
\label{f:PWN-aldist}
\end{figure}
%------------------------------------------------------------
           
The spatial distributions of $F$ and $\alpha_{\nu}$ in Figs.~8 and 9
appear to have similar shapes and may suggest a harder spectrum from 
the dimmer outer areas of the PWN. The data are, however, rather uncertain 
and a constant spectral index of $\sim 1$ seems to be compatible with 
all errorbars in Fig.~9.   
To check that more thoroughly we analyzed  the $\alpha_{\nu}$--$F$  
distribution  presented in Fig.~10 which also includes both jet areas. 
A linear regression fit to $\alpha_{\nu}$ versus log$(F)$ using 
the method described  in this Sect. above yields 
 \begin{equation}
  \alpha_{\nu} = 19.44 + 1.35~{\rm log}(F)  
 \end{equation} 
This is shown by a solid line in Fig. 10,
and indicates that brighter structures of the PWN have a steeper spectral 
index. The gray area shows the 1$\sigma$ uncertainty of the fit. 
The range of the line slopes within this area 
is $1.35^{+1.02}_{-0.98}$. A zero slope, and thus no correlation, is allowed 
only if we increase the uncertainty area to 1.4$\sigma$. This value 
can be considered as an overall significance of a possible 
 $\alpha_{\nu}$-log$(F)$ correlation.
We also checked the spectral index-flux correlation using the spectral 
fit parameter $F_{\nu_0}$  as well as the measured flux 
value $F_{\nu}$ in the F547M band (see Table~\ref{t:PWN-phot})    
instead of the derived $F$ and got only a slightly tighter correlation.
%-----------------------figure10------------------------------
\begin{figure}[t]
\begin{center}
\includegraphics[width=85mm, angle=270, clip]{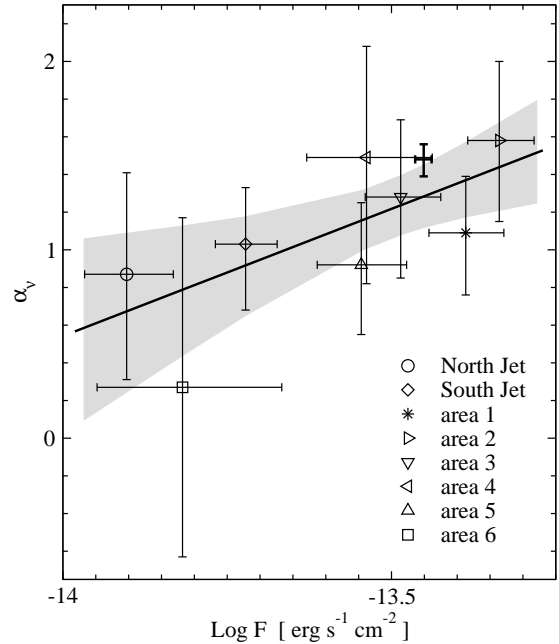}
\end{center}
\caption{Spectral index ${\rm \alpha_{\nu}}$ versus
optical flux, log$(F)$, in the $(3.8-8.9)\times10^{14}$ Hz range
for the different areas of the PWN marked according to
Fig.~\ref{f:PWNstady}.
The thick solid line shows the best fit linear model given by
Eq. (4). The point marked by bold errorbars shows the parameters for
the full PWN rescaled to a 7-pixel aperture (cf. Fig. 8). The gray area
shows the 1$\sigma$ uncertainty area for possible linear fits to the data 
points using the method described in Sect. 2.5. A zero slope correlation
is excluded at the 1.4$\sigma$ level.
}
\label{f:PWN-corr}
\end{figure}
%------------------------------------------------------------  

Therefore, deeper observations of the PWN are needed to probe a
correlation between the brightness and the spectral index 
which is indeed only marginally indicated by the current 
optical data. A study of the index-flux distribution in X-rays  
would also be useful for the PWN of \psr, as has recently been done for the 
Crab PWN (\cite{Mori04}). The study of Mori et al. shows that for brightnesses 
above $\sim$0.7~counts~s$^{-1}$~arcsec$^{-2}$ 
there is a hint that (cf. their Fig.~3) the spectral index of the 
torus region increases with the surface brightness, as marginally 
indicated also in our case for \psr. However, including brightnesses down 
to $\sim$0.4~counts~s$^{-1}$~arcsec$^{-2}$,  
the spectral index -- surface brightness distribution appears flat. 
The indices of the Crab jet are generally 
smaller compared to those of peripheral PWN regions although their 
surface brightnesses are comparable (Mori et al.~2004).  
In the case of \psr\ the correlation may be enhanced  
by a larger brightness asymmetry of the PWN. We note that such 
a correlation, as well as a flat index versus flux distribution,   
contradicts simple expectations from synchrotron cooling of relativistic 
particles which suggest a softening of the underlying electron 
spectrum toward the PWN boundary. In this picture, the fainter outer regions 
of the nebula would emit softer spectra. However, this simple picture  
does not work even for the much better studied Crab, 
where the PWN torus does not change its size significantly from  
radio to hard X-rays, whereas the respective cooling times  
differ by many orders of a magnitude implying much larger extents 
in the radio and optical than in X-rays. The same appears to be true 
for \psr\ (\cite{Car00}). This is not yet explained, neither by the 
classical isotropic pulsar-wind model of a PWN by Kennel \& Coroniti (1984),
nor by modern MHD versions of it (\cite{bk02}; \cite{kl04}; 
\cite{zab04}) despite the fact that these also include anisotropy of 
the wind along the pulsar rotation axis, and can qualitatively explain 
the observed torus-jet structure invoking a complicated axisymmetric 
picture of the wind termination shock in the internal region 
of the PWN.

\subsection{Proper motion and astrometry of \psr.}

The position of \psr\ is defined on the HST PC chip frames with an accuracy
of better than 0.17 PC pixels which corresponds to 0\farcs0077. 
This is only a factor of $\sim 1.5$ larger than the yearly proper 
motion value reported by Manchester et al. (1993b), see Table~\ref{t:param}.
This allows a direct estimate of the proper motion of the pulsar using accurate 
superposition of the F555W and F547M images taken at 
epochs separated by 4 years (see Sect.~2.3). 
We used the positions of 9 reference stars to construct 
the coordinate transformation between the two images with the IRAF routines 
{\sf geomap} and {\sf gregister}. 
The {\sl rms\/} errors of the transformation fit were   
0.078 and $4.6\times10^{-4}$ of the PC pixel size in RA and DEC,  
respectively, with residuals being  $\le 0.156$ pixels  in RA and 
 $8.1\times10^{-4}$ pixels in DEC.
Using {\sf imcentroid} for measuring the coordinates of the pulsar 
we find a shift of $0.431\pm0.203$ pixels between 
its positions for a time difference of 3.995 years, where the error 
accounts for the centroid and transformation uncertainties. 
This corresponds to a proper motion $\mu=4.9\pm2.3$ mas y$^{-1}$  
in the South-East direction at a position 
angle of 108\fdg 7$ \pm 32\fdg 9$ (along the southern jet). 
The significance of this result is low and can be considered  
only as an attempt to make a first direct measurement of the proper motion. 
Based on the displacement between the pulsar optical position and 
the center of the PWN, as seen in radio,  
Manchester et al. (1993b) argued for a similar value of the proper 
motion but in the South-West direction  (in the plane of the torus).        
We note that the proper motion of the Crab 
pulsar is aligned with the symmetry axis of the inner Crab nebula, as 
defined by the direction of the X-ray jet discovered by 
ROSAT (\cite{Car99}), and that a similar situation applies 
to the Vela pulsar (\cite{Luca00}; \cite{Car01}; \cite{Dod03}). 
If our estimates are close to reality,  
we have the intriguing situation that all these three young pulsars 
appear to move along the jet axis. 
A difference is, however, that while the Crab and
Vela pulsars both have transverse velocities of $\sim 130\kms$, our
results for \psr\ indicate
a higher transverse velocity 1190$\pm560\kms$, assuming a distance 
to the LMC of 51 kpc (Panagia 2004). 
A third
epoch of HST imaging to confirm the large value and direction for the 
transverse velocity is clearly needed to establish this result 
at a higher significance level. Based on our proper motion estimates 
a level of $\ga3\sigma$ can be achieved starting from the beginning of 2005.

We also determined the coordinates of the 
pulsar in the F547M image using seven GSC-II stars visible within 
the PC chip frame and the IRAF routines {\sf ccmap}, {\sf cctran}, 
and {\sf ccsetwcs}. The formal {\sf rms} errors of the astrometric  
fit are 0\farcs423 and 0\farcs472 in the RA and DEC, respectively. 
Combined with the nominal GSC-II catalog accuracy 
of 0\farcs5, this gives an  accuracy of
the position of the pulsar of 0\farcs655 and 0\farcs688  
in RA and DEC, respectively.
In Table~\ref{t:psr-coord} we compare our astrometry  
with previous results. We note the good agreement between our measurement
and the latest Chandra result (Kaaret et al. 2001). 
%%%%%%%%%%%%%%%%%%%%%%%%%%%%%%%% Table 7 astrometry of psr coords %%%%%%%%%%%%%%%%%%%%%%%%%  
\begin{table}[t]
\caption{Coordinates of \psr.}
\label{t:psr-coord} 
\begin{tabular}{lll}
\hline
\hline
Range    & R.A.(J2000) & Dec. (J2000) \\ 
         & {05\h 40\m} + &  {-69\degs 19\amin} + \\ 
\hline 
X-ray$^a$   & {11\fss 0 $\pm$ 0\fss 7}     & {57\farcs 4 $\pm$ 2\farcs0} \\  
Optical$^b$ & {10\fss 99 $\pm$ 0\fss 18}   & {55\farcs 1 $\pm$ 0\farcs5} \\
Radio$^c$   & {11\fss 1}                   & {57\farcs 5} \\
X-ray$^d$   & {11\fss 221 $\pm$ 0\fss 132}   & {54\farcs 98 $\pm$ 0\farcs7} \\
Optical$^e$ & {11\fss 173 $\pm$ 0\fss 121} & {54\farcs41 $\pm$ 0\farcs7} \\ 
\hline
\end{tabular}\\
$^a$~Seward et al. (1984); $^b$~Caraveo et al. (1992); \\
$^c$~Manchester et al. (1993a); $^d$~Kaaret et al. (2001); \\
$^e$~this work 
\end{table} 
%%%%%%%%%%%%%%%%%%%%%%%%%%%%%%%%%%%%%%%%%%%%%%%%%%%%%%%%%%%%%%%% 

\subsection{X-ray spectrum}

\subsubsection{Interstellar absorption}
The new optical results allow us to update the multiwavelength picture 
of the emission of \psr. We will discuss this in greater detail
in Sect. 3. Before establishing the overall spectrum, one first
needs to accurately correct for absorption and scattering by the
interstellar gas and dust. In a recent analysis of time-resolved Chandra 
data, Kaaret et al. (2001) argue that the pulsed emission of \psr\ can be 
approximated by a power-law $F_{E}=F_{E_0}(E/E_0)^{-\alpha_{E}}$
with $\alpha_{E} = 0.83\pm0.13$ within the photon energy range $E = 0.6-10.0$ keV range,
provided that the foreground photoelectric absorption is caused by gas with 
Milky Way (MW) abundances and a column density 
of $N_{\rm HI} = 4.6\times10^{21}$~cm$^{-2}$. A similar number 
($N_{\rm HI} = 4.0^{+0.6}_{-0.4}\times10^{21}$~cm$^{-2}$) was estimated
by Finley et al. (1993) analyzing ROSAT X-ray data.

However, the use of MW abundances is obviously a simplification for \psr.
As a matter of fact, only a fraction of the photoelectric absorption is
likely to occur in the Milky Way. The recent Parkes 21~cm multibeam survey 
of the LMC (\cite{stav03}) shows that the MW contribution to the
column density in the direction to \psr\ is 
just $N_{\rm HI} \approx 0.6\times10^{21}$~cm$^{-2}$.
This survey also shows that the maximum value of $N_{\rm HI}$ in the LMC
is $\sim 5.6\times10^{21}$~cm$^{-2}$, and that this occurs close to the 
position of \psr. This is consistent with the hydrogen column density found 
by fitting the wings of the Ly${\alpha}$ absorption profile for the neighboring 
LMC star Sk~$-$69~265, $N_{\rm HI} = (5\pm0.5) \times10^{21}$~cm$^{-2}$
(\cite{gor03}). It is quite likely that \psr\ could have a similarly high
column density, especially since its dispersion measure (see Table~1)
is $\sim 50$\% higher than for any other pulsar in the LMC (\cite{crawf01}).
A rough estimate of the LMC part of $N_{\rm HI}$ for \psr\ can be obtained
from scaling of the estimated column density for SNR~1987A and its neighboring
star, ``Star 2''. Michael et al. (2002) used LMC abundances to derive
$N_{\rm HI}({\rm LMC}) = 2.5^{+0.4}_{-0.3}\times 10^{21}$ cm$^{-2}$ for
SNR~1987A, and Scuderi et al. (1996) 
obtained $N_{\rm HI}({\rm LMC}) = 3.4^{+1.0}_{-0.9}\times 10^{21}$ cm$^{-2}$ 
for Star 2 allowing for a foreground MW contribution 
of $0.6\times10^{21}$~cm$^{-2}$. We 
adopt $N_{\rm HI}({\rm LMC}) = 2.9^{+1.0}_{-0.9}\times 10^{21}$ cm$^{-2}$ for
SNR~1987A and Star 2. Assuming that SNR~1987A and \psr\ suffer similar 
amounts of LMC absorption in proportion to the LMC 21~cm emission 
(\cite{stav03}) along their respective lines of sight, we 
obtain $N_{\rm HI}^{0540} \approx N_{\rm HI}^{87A}~ 
{\rm HI}^{0540}/{\rm HI}^{87A} \approx
5.5^{+1.9}_{-1.7} \times 10^{21}$ cm$^{-2}$, 
where ${\rm HI}^{0540}= 32.00$~Jy/beam and ${\rm HI}^{87A} = 16.75$~Jy/beam 
are the line flux densities toward the pulsar and SNR 1987A, respectively, 
according to the Parkes survey data 
base\footnote{www.atnf.csiro.au/research/multibeam/release/}. 
The assumptions used to obtain this result 
for $N_{\rm HI}^{0540}$ are of course uncertain, but the result points in 
the same direction as the estimates from the 21 cm emission, Sk~$-$69~265 
and the pulsar dispersion measure discussed above, i.e., the column 
density for \psr\ is high. Assuming that the 
21 cm emission at the position of \psr\ marks an upper limit to
its $N_{\rm HI}({\rm LMC})$, we can limit the range 
to $N_{\rm HI}^{0540} = (4.6\pm1.0)\times 10^{21}$ cm$^{-2}$. This is 
similar to what was used by Kaaret et al. (2001), but with the important 
difference that the X-ray absorption is not mainly Galactic, 
but arises in the LMC.

To illustrate the effect of LMC abundances we show in Fig. 11 the ratio of 
photoelectric cross sections (per hydrogen atom) in the LMC and MW for the 
energy range 0.1--10 keV. We will refer to this ratio as $f(E)$. The drop 
in $f(E)$ at $E > 0.28$ keV (the K-shell edge of carbon) just reflects the 
lower metal content in the LMC compared to the Galaxy. For the MW we have 
used the abundances in Morrison \& McCammon (1983, henceforth MM83), and for 
LMC we have adopted the abundances of He, C, N, O, Mg, Si and Fe from Korn et 
al. (2002). We have also included the elements Ne, Na, Al, S, Ar and Ca 
for which we have assumed that the LMC abundances are 0.4, 0.4, 0.5, 0.4, 
0.5 and 0.5 times the solar values in MM83, respectively. The exact numbers 
for these elements are not important for our analysis 
since the absorption is dominated by C, N, O and Fe in the 
energy range we are most interested in. We assume that the interstellar
gas is neutral, and we disregard dust. Photoionization cross sections were
taken from the code used in Lundqvist \& Fransson (1996) with further
updates for Na, Mg, Al, Ar, \& Ca using the TOPbase archive (Cunto \& Mendoza
1993), as well as for He (Samson et al 1994; Pont \& Shakeshaft 1995).
We have tested this code against the results of MM83 for solar abundances,
and the cross sections agree to the same level of accuracy as the
recent cross sections compiled by Wilms et al. (2000). The comparison
against MM83 is relevant as Kaaret et al. (2001) did their analysis using
XSPEC Ver. 10.0 which uses data fully compatible with MM83. Figure 11
shows that LMC abundances strongly suppress the photoelectric absorption,
and that C and O are particularly important constituents at the energies
for which Kaaret et al. (2001) claim photoelectric absorption is most
important for the observed pulsar spectrum, i.e., at $\lsim 1$ keV.

%-----------------------figure11------------------------------
\begin{figure}[tbh]
\begin{center}
\includegraphics[width=60mm,clip,angle=90]{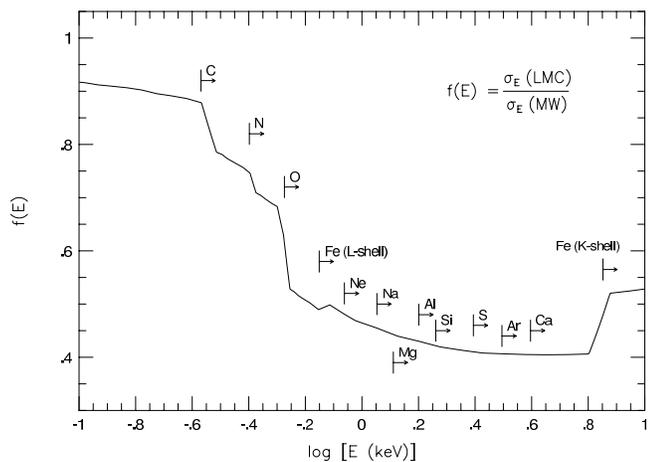}
\end{center}
\caption{Ratio of photoelectric absorption per hydrogen atom in the LMC
and Milky Way for the energy range 0.1--10 keV.
Arrows mark at which energy each element starts to contribute to the 
photoelectric absorption. The absorbing gas is assumed to be neutral. The
gradual changes of $f(E)$ across absorption edges are due to
moderate zoning of the photon energy in the code used for the cross section
calculations. See text for further details. 
} \label{f:OptDepth}
\end{figure}
%------------------------------------------------------------      

We have used the function $f(E)$ in combination with the results of MM83 
to see how LMC abundances may affect conclusions about the derived 
spectrum of \psr. If we assume that the pulsar emits a pulsed power-law 
spectrum with slope $\alpha_{E} = -0.83$ in the range $0.6-10.0$ keV, as 
argued for by Kaaret et al. (2001) and marked in Fig. 12 as a straight 
dotted line, the attenuated spectrum should look like the solid line in 
Fig. 12 after passage through a column density of neutral hydrogen in the 
Milky Way with a value of $N_{\rm HI}({\rm MW}) = 4.6\times 10^{21}$ cm$^{-2}$.
If we disregard possible effects of an accurate treatment for the response
matrix of Chandra, we can deabsorb this spectrum with a more
likely composition for the photoelectrically absorbing gas. For this
we have chosen $N_{\rm HI}({\rm MW}) = 0.6\times 10^{21}$ cm$^{-2}$
and $N_{\rm HI}({\rm LMC}) = 5.0\times 10^{21}$ cm$^{-2}$. The
deabsorbed spectrum is marked by the upper dashed line in Fig. 12. 
At the lower energy limit of the fit by Kaaret et al. (2001), i.e., 
at 0.6 keV, the deabsorbed spectrum undershoots by a factor of $\sim 2.9$
compared to the power-law, but on the other hand overshoots by orders
of magnitude at energies below the K-shell edge of carbon. The latter can
be fixed by just lowering $N_{\rm HI}({\rm LMC})$ 
to $\lsim 4.3\times 10^{21}$ cm$^{-2}$, i.e., still consistent with the likely 
range argued for earlier in this section. The spectrum would in that case
undershoot by a factor of $\gsim 3.8$ at 0.6 keV compared to the power-law. 
The assumption of a power-law spectrum at energies below a few keV, where
the photoelectric absorption sets in, is of course uncertain.
Our results could indicate that the intrinsic spectrum is not a power-law, but 
actually falls below the power-law at 0.6 keV. However, before jumping 
to such a conclusion we need to check another possible source of X-ray 
absorption, namely the supernova ejecta.
%-----------------------figure12------------------------------
\begin{figure}[tbh]
\begin{center}
\includegraphics[width=60mm,clip,angle=90]{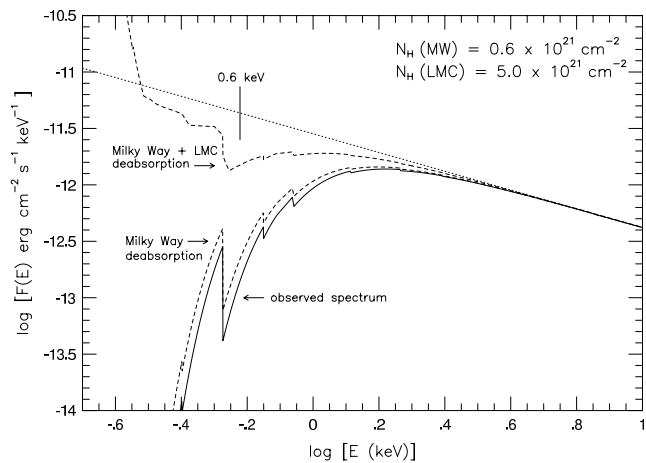}
\end{center}
\caption{Soft X-ray spectrum of the pulsed emission from \psr. The dotted
line shows the intrinsic power-law spectrum argued for by Kaaret et al. (2001), 
and the solid line shows the attenuated spectrum after passing through a 
column of gas with MW 
abundances, $N_{\rm HI}({\rm MW}) = 4.6\times 10^{21}$ cm$^{-2}$. The 
lower dashed line shows deabsorption using a more realistic value for the MW
contribution, $N_{\rm HI}({\rm MW}) = 0.6\times 10^{21}$ cm$^{-2}$, and the
upper dashed line shows further deabsorption with the 
value $N_{\rm HI}({\rm LMC}) = 5.0\times 10^{21}$ cm$^{-2}$ for LMC, 
which is close to the maximum value found for the LMC as a whole
according to Staveley-Smith et al. (2003). The vertical bar at 0.6 keV
marks the energy above which Kaaret et al. (2001) fitted their power-law
to the deabsorbed spectrum.
} \label{f:LMCMW}
\end{figure}
%------------------------------------------------------------      

%-----------------------figure13------------------------------
\begin{figure}[tbh]
\begin{center}
\includegraphics[width=60mm,clip,angle=90]{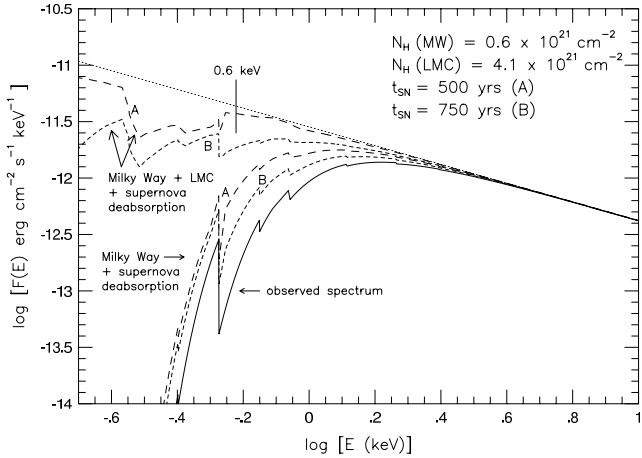}
\end{center}
\caption{Same as Fig.~12, but with a lower value for the LMC
deabsorption, $N_{\rm HI}({\rm LMC}) = 4.1\times 10^{21}$ cm$^{-2}$.
Deabsorption due to supernova ejecta has also been included,
assuming the same structure as in SN~1987A. 
The structure of SN~1987A on day 118 (Blinnikov et al. 2000) was
expanded homologously until 500 years (model ``A'', long-dashed) or 750 years 
(model ``B'', short-dashed) after the explosion. Note that the age-parameter
in these models do not necessarily relate to the same age for
SNR~0540-69.3 as it depends on the explosion energy, the envelope mass
and the density structure. See text for further details.
} \label{f:LMCMWSN}
\end{figure}
%------------------------------------------------------------      

\subsubsection{Supernova ejecta absorption}
Observations of SNR~0540-69.3 show that the remnant is oxygen-rich
(Kirshner et al. 1989; Serafimovich et al. 2004), which means that
the progenitor star probably was massive. Kirshner et al. (1989)
suggest it to have had a mass similar to that of the SN~1987A 
progenitor, i.e., around $20 \Msun$, and the results of Serafimovich et al.
(2004) show that a likely range is $15-22 \Msun$. To check the 
importance of supernova ejecta for possible X-ray absorption,
the well-studied SN 1987A therefore serves as a good template.
We have used the mixed model 14E1 displayed in Figs. 2 and 6 in Blinnikov et
al. (2000), which has $\approx 2 \Msun$ of pure oxygen spread out over 
the innermost $\sim 5 \Msun$ of supernova ejecta. We assume that the
post-explosion structure expands homologously. At an age 
of 1000 years after explosion, the hydrogen column density through the
ejecta, from the center, is 
only $N_{\rm HI}({\rm SN}) \approx 1.1\times 10^{19}$ cm$^{-2}$, i.e., far 
less than even the MW contribution in the direction toward LMC. However, 
the very high metal content in the supernova ejecta, and its concentration 
toward the center of the ejecta,
boosts the X-ray absorption at energies above 300 eV. Above
the K-shell ionization edge of oxygen, i.e. at 600 eV, the cross section per
hydrogen atom is $\sim 40$ times higher than for the MW absorption in
MM83. Because $N_{\rm HI}({\rm SN})$ scales with the time since explosion 
as $\propto t^{-2}$, a less evolved supernova remnant can make a significant
contribution to the X-ray absorption of emission from the pulsar and
its PWN.

The importance of the supernova ejecta is illustrated in Fig. 13. Here we have
used the same value for $N_{\rm HI}({\rm MW})$ as in Fig. 12, but
lowered the LMC contribution 
to $N_{\rm HI}({\rm LMC}) = 4.1\times 10^{21}$ cm$^{-2}$.
We have also included deabsorption due to supernova ejecta, assuming 
a homologously expanding SN~1987A at an age of 500 years (model ``A'', 
long-dashed) and 750 years (model ``B'', short-dashed). 
The increased oxygen column density compared to that in Fig. 12 makes it
possible to retrieve the initial power-law spectrum above 0.6 keV
for a supernova similar to SN~1987A at an age of 500 years, but it is
also clear that a remnant with an age closer to the spin-down age of \psr\ 
will not contribute significantly to the X-ray absorption. Although the 
ejecta of SNR~0540-69.3 could expand more slowly than those of SN 1987A, 
could contain more oxygen (and/or having it more concentrated to the center), 
could have a clumpy and asymmetric structure (as is indicated for SN~1987A, 
Wang et al. 2002), or could have a lower age than the pulsar spin-down age, 
it seems that we have to stretch the parameters to claim that the X-ray 
absorption along the line of sight to the pulsar is largely affected by
the supernova ejecta. A possible way to test this is to check the spatial
variations of the X-ray absorption over the larger PWN. The low metal
content of LMC, and the low MW foreground absorption, make such tests 
sensitive to any supernova ejecta contribution. X-ray spectral fits to data 
obtained with Chandra of the Crab pulsar have recently highlighted the 
importance of the abundance factor (\cite {Will01}; \cite {w03}). 
Using various abundances 
it was found that the line of sight to the Crab is significantly 
underabundant in oxygen.

To summarize this section, it is evident that the X-ray spectral analysis 
of \psr\ needs a revision. Contrary to previous assumptions, 
the metal abundance of 
the X-ray absorbing gas must clearly be sub-solar, unless the supernova ejecta 
contribute significantly. The latter, however, appears to be less likely. 
While a full analysis of the X-ray spectrum, i.e., a detailed reduction
of the Chandra data including various abundance combinations in XSPEC, 
is beyond the scope of this paper, we have argued that 
the power-law spectrum, which seems to be appropriate to use at energies 
above $\sim 1$ keV, may experience a depression below $\sim 1$ keV (cf.
Sect. 3). This can be tested by how the X-ray spectrum connects to that in
the optical. We will discuss that in Sect 3.

\subsection{Extinction toward \psr }
As for the column density toward \psr, the foreground optical extinction
in the MW is well established. From the study of
Staveley-Smith et al. (2003), the color excess in the direction 
toward \psr\ is $E(B-V)_{\rm MW}=0.06$, and this agrees with the value
found for SN~1987A (\cite{scu96}) and that used by Gordon et al. (2003).
The value used for the parameter $R$ in these works is the standard 
value 3.1.

The extinction parameters $E(B-V)$ and $R(V)$ for LMC along the line of 
sight to SNR~0540-69.3 and the pulsar have not yet been investigated in 
detail in the same way as for SN~1987A (\cite{scu96}). 
However, in Serafimovich et al. (2004) we studied SNR~0540-69.3 and H~II
regions close to it. The reduced spectra were analyzed using the total 
value $E(B-V)_{\rm TOT} = 0.20$ and $R(V) = 3.1$, i.e., the same numbers
we used in Sect. 2, and we found that the 
H$\beta$/H$\gamma$/H$\delta$ line ratios are in good agreement with 
Case~B recombination theory (Baker \& Menzel 
1938; Hummer \& Storey 1987). Allowing for higher extinction would
boost H$\gamma$ and H$\beta$ relative to H$\beta$ causing a disagreement
with the Case~B theory, which for these lines normally explains
the observations of supernova remnants well (e.g., Fesen \& Hurford 1996). 

Looking at the projected pulsar neighborhood, Gordon et al. (2003) obtain
the average value $R(V)_{\rm LMC}=2.76\pm0.09$ for the LMC2 supershell, 
and it seems reasonable that this could be used also for \psr.
Out of the eight stars forming this average, six of them can accommodate 
the standard value of 3.1 within $1\sigma$. The spread in $E(B-V)_{\rm LMC}$ 
ranges between 0.12--0.24 (including $1\sigma$ errors), so the values we
have used for \psr\ and its PWN in Sect. 2.4 and 2.5 seem reasonable also 
from this comparison. However, to check the effect of a different extinction 
curve on $E(B-V)$, and still being compatible with Case~B line ratios for 
SNR~0540-69.3, we have compared $k_{ext}$ (cf. Sect. 2.4) for the extinction 
used in Sect. 2.4 (we call that case ``C1'') to a case (called ``C2'')
with $E(B-V)_{\rm LMC}=0.14$ ($R(V) = 2.76$, Gordon et al. 2003) 
and $E(B-V)_{\rm MW} = 0.06$ ($R(V) = 3.1$, Cardelli et al. 1989). We 
formed the ratio $g(\lambda) = k_{ext}(\lambda)/k_{ext}({\rm H}\beta)$ for
both cases, and found 
that $\mid g(\lambda)_{C1} - g(\lambda)_{C2} \mid / g(\lambda)_{C1}$ 
does not exceed 2\% within the interval 2620--8480~\AA. Only at the very 
blue end of the FOS spectrum of Hill et al.  (1997), i.e., 
at $\approx 2500$~\AA, does the ratio approach 5\%. This justifies 
the use of $E(B-V)_{\rm LMC} = 0.14$ (assuming $E(B-V)_{\rm MW} = 0.06$)
regardless of whether we choose to use extinction combinations like C1 or 
C2. The absolute flux level of the dereddened spectrum of course depends 
on the exact value of $R(V)$ being used. Further direct extinction studies 
of the pulsar and its neighborhood in the UV and optical bands are 
needed to pin down the detailed extinction corrections. 
It seems, however, that the steep spectral
slopes we obtain for the pulsar and its PWN in the optical in Sect. 2.4 and
2.5 cannot be corrected by some extreme reddening corrections as this is 
neither justified by our observations of the supernova remnant nor by the
study by Gordon et al. for supposedly neighboring objects.

\section{Discussion}

\subsection{Multiwavelength spectrum of \psr}

%-----------------------figure14------------------------------
\begin{figure*}[thb]
\begin{center}
\includegraphics[height=180mm, clip, angle=270]{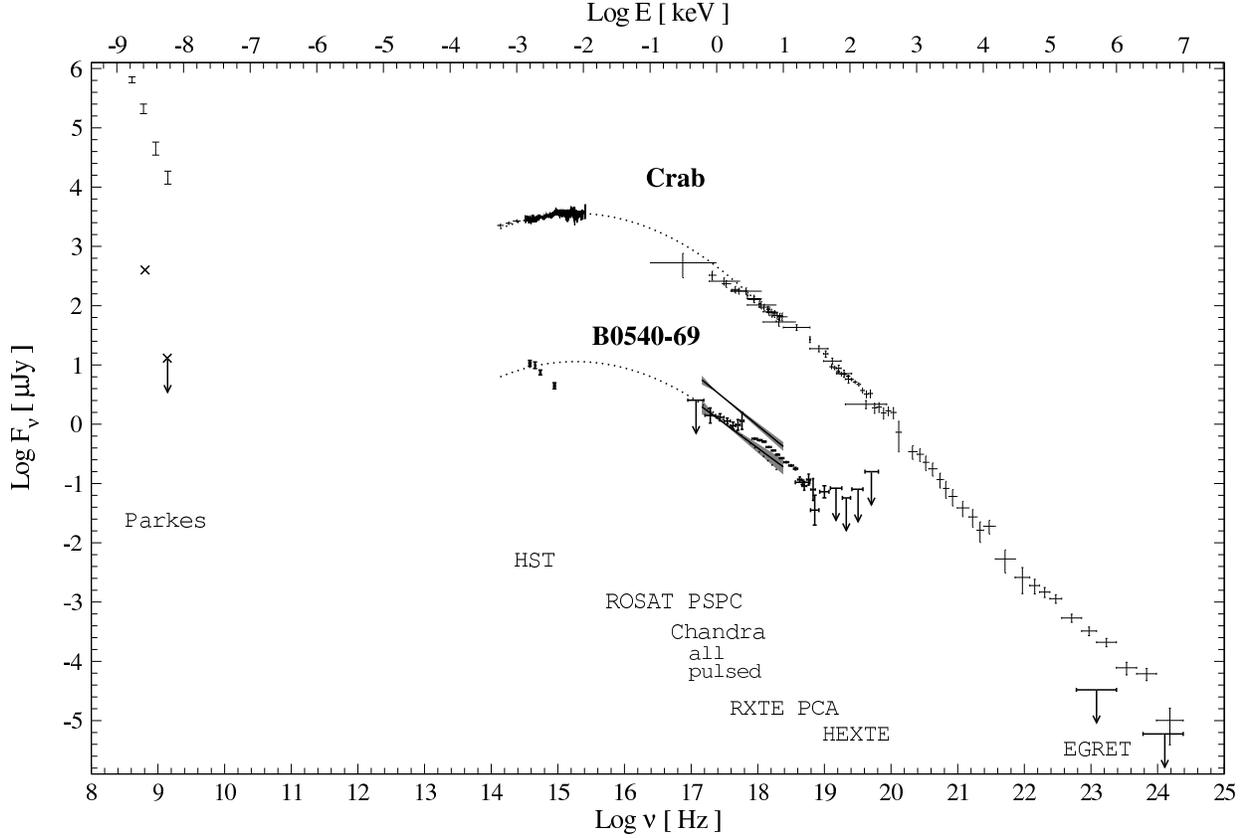}
\end{center}  
\caption {Multiwavelength unabsorbed spectrum of \psr. The data
were obtained with the different telescopes and instruments marked in the plot.
The optical data are from this paper. Phase-averaged (upper polygon) and pulsed 
(lower polygon) X-ray emission spectra with their uncertainties 
are shown for the 0.6 -- 10 keV range (Kaaret et al. 2001). 
The pulsed emission spectra obtained with ROSAT and RXTE 
are from de Plaa et al. (2003), and the radio and $\gamma$-ray data are 
from Manchester et al.~(1993a) and Thomson et al. (1994), respectively.
All upper limits are defined at $3\sigma$ level.  
The unabsorbed spectrum of the total pulsed emission of the Crab pulsar   
is shown for comparison, where the high energy and optical data are from 
Kuiper et al. (2001) and Sollerman (2003), respectively.    
The dotted line for the Crab pulsar indicates a possible connection 
between the optical/X-ray regions through the EUV. The same
line (shifted to overlap with the soft X-ray spectrum) is overlaid on the optical/X-ray 
region of \psr. Note how the blue band fluxes of 
\psr\ undershoot by a factor of $\sim$3 compared to this 
overlay while it fits well in the X-ray band.}   
\label{f:totalspec} 
\end{figure*}
%------------------------------------------------------------

To connect the optical pulsar emission to the emission at other 
wavelengths, we have compiled results for the radio (\cite{Man93a}), 
X-ray (\cite {Kaa01}; \cite{Plaa03}) and $\gamma$-ray (\cite{Thom94})
spectral regions. The unabsorbed spectrum is displayed in 
Fig.~\ref{f:totalspec}. In comparison with previous
compilations (e.g., Hirayama et al. 2002; \cite{Plaa03}) the accuracy is 
significantly improved due to the new high spatial resolution data 
obtained in the optical with HST and in X-rays with Chandra. 
This reveals new features in the spectrum of the pulsar.
   
%-----------------------table8------------------------------
\begin{table*}[hbt]
\caption{Comparison of the optical and X-ray  spectral indices 
($ {\rm \alpha^{O}_{\nu}}$, ${\rm \alpha^{X}_{\nu}}$),   
luminosities (${L^{\rm O}}$, $ {L^{\rm X}} $),  efficiencies (${ \rm \eta ^{O} } $, ${ \rm \eta ^{X} } $), 
and a weighted ratio ($ {L^{\rm O}\Delta E^{\rm X}/L^{\rm X}\Delta E^{\rm O}} $) of the two young pulsars, Crab and \psr, 
and the older Vela pulsar. Here $\Delta E^{\rm O}$ and $\Delta E^{\rm X}$ are
the energy intervals in the optical$^a$ and X-rays$^b$, respectively, used for
the frequency integration to obtain $L^{\rm O}$ and $L^{\rm X}$.
Information on the period $P$, dynamical age $ {\rm \tau } $, spindown 
luminosity $\dot E$, and distance $d$ for each pulsar is included.}  
\label{t:psr-lum}
\begin{tabular}{lccccccccccc}
\hline\hline 
PSR & $P$ & ${\rm \tau}$ & ${\dot E}$ & $d$ & ${\rm \alpha^{O}_{\nu}}$  
& ${L^{\rm O,a}}$ & ${\rm \eta ^{O}}$ & ${\alpha^{\rm X}_{\nu}}$ & ${L^{\rm X,b}}$  
& ${ \rm \eta ^{X}}$ & ${L^{\rm O}\Delta E^{\rm X} \over L^{\rm X}\Delta E^{\rm O}}$  \\   
  & ms & kyr & ${\rm 10^{37}~erg~s^{-1}}$ & kpc &   & ${\rm 10^{33}~erg~s^{-1}} $  & ${\rm  10^{-5}  } $  &   
  & ${\rm 10^{36}~erg~s^{-1}} $   & ${\rm  10^{-3}  } $  &   \\
\hline  
Crab &  33.49  & 1.24 & 46  & 2 & -0.21$^c$  & 7.9$^c$  & 1.7 & 0.87$^d$  & 1.0$^d$ & 2.17 & 34.9  \widerul\\
0540  &  50.2  & 1.66 & 15 & 51 &  1.07$^e$  & 10.9$^e$  &  7.3 & 0.83$^d$  & 3.13$^d$ & 21  & 15.5  \widerul\\    
Vela  &  89  & 11 &  0.69 & 0.29 & 0.25$^f$   &  ${\rm 7.0^f\times 10^{-5}}$ & ${\rm 1.0\times 10^{-3}}$  & 0.5$^g$  & 
${\rm 2.3^g\times 10^{-5}}$ & ${\rm 3.4 \times10^{-3}}$  & 13.5 \widerul\\   
\hline
\end{tabular}
\begin{tabular}{ll}
$^a$~For the optical range 1.57--3.68 eV. & $^e$~This paper.\\ 
$^b$~For the X-ray range 0.6--10 keV. & $^f$~Using data from Shibanov et al. (2003). \\
$^c$~Using data from Sollerman (2003) for 1.57--3.68 eV. & $^g$~Pavlov et al.~(2001a); ${L^{\rm X}}$ rescaled to the 0.6--10 keV range \\
$^d$~Kaaret et al. (2001); $ {L^{\rm X}}$ rescaled to the 0.6--10 keV range. & and $d=0.29$~kpc.\\
\end{tabular}
\end{table*}

The optical and X-ray  parts of the spectrum  can be fitted with  power-laws  
(for specific assumptions about the intervening extinction and photoelectric 
absorption discussed in Sect.~2.7 and 2.8), which would suggest a non-thermal 
nature of the emission in both domains, likely to be formed in the 
magnetosphere of the rotating neutron star. However, the connection between
the emission in X-rays and in the optical seems far from trivial, especially 
if the slope for the optical spectrum derived from the archival 
HST/WFPC2/F336W data is correct (spectral indices and other emission 
parameters for \psr, as well as for the Crab and Vela pulsars,  
in the optical and X-rays are presented in Table~8). 
The data for \psr\ suggest at least two spectral breaks 
between the optical and X-ray spectral bands. For a comparison, 
Fig.~\ref{f:totalspec} also shows the multiwavelength spectrum of the total 
pulsed emission from the Crab pulsar (Kuiper et al. 2001; Sollerman 2003). 
For the Crab, it seems that a smooth turn-over can be possible (dotted line) 
between the X-ray band and the optical. This is in contrast to spectra 
of the middle-aged pulsars Vela and PSR B0656+14, whose optical fluxes are 
generally compatible with the low-frequency extrapolation of a power-law 
spectral tail for $E\ga 1-2$~ keV (Koptsevich et al.~2001; 
Shibanov et al.~2003).     

In Sect. 2.7 we noted that \psr\ could have a non-powerlaw spectrum
below $\sim 1-2$ keV, and inspired by this we tried to use the
shape of the Crab optical/X-ray turn-over to fit the spectrum of \psr. This, 
however, fails for the WFPC2/F336W band where the flux falls below such
a fit. As we pointed out in Sect. 2.8, the depression in U is unlikely to be 
caused by insufficient dereddening; to reach the Crab pulsar spectral 
slope ($\alpha_{\nu}\approx -0.2$, cf. Table 8) one has to 
apply $E(B-V)\approx 0.55$, 
which is at least twice as high as the most likely value (cf. Sect.~2.8). 
Does this mean that \psr\ experiences
a spectral dip in the F336W band? While future deep and well-calibrated data
in U and UV should reveal this, we note that such an
explanation is not farfetched. As a matter of fact, the  broad-band 
optical spectra of middle-aged pulsars (Vela, PSR B0656+14, and Geminga, 
cf. \cite{shib03}) do show a dip in the U and B bands, which could
indicate a multicomponent continuum spectrum, or the presence of unresolved 
emission/absorption features, possibly related to  electron/ion cyclotron 
lines originating in the magnetospheres of the neutron stars 
 (\cite{mig98};  \cite{jac99}).
If the depression in U is of more general character, the multiwavelength
spectrum of \psr\ suggests a double break ``knee'' in the spectral region 
between the optical and soft X-ray bands. Observations in the UV and 
reanalysis of the Chandra X-ray data with accurate corrections for 
extinction and photoelectric absorption, as discussed in Sect. 2.7,
will help us constrain the position of the breaks and to understand whether 
they are located just blueward of the U band and below 0.6 keV, or occur
in the EUV range. 
 
Taking into account the difference in distance to the Crab pulsar ($\sim$2~kpc) 
and \psr\ ($\sim$51 kpc), as well as the spectral energy distributions for 
both pulsars (see Fig.~14), we note that the overall intrinsic flux 
from \psr\ is almost as high as that from the Crab pulsar in the radio 
(while its slope is possibly steeper in this range), but that it is $\sim 1.4$
and $\sim 3$ times higher in the optical and X-ray ranges, respectively. This 
is also shown in Table~8. At the same time, the spindown luminosity, $\dot{E}$,
of \psr\ is $\sim 3$ times lower than for the Crab. Therefore, the 
efficiency of producing nonthermal optical and X-ray photons in the
magnetosphere of the rotating neutron star from its spindown 
power, $\eta = L/\dot{E}$, is a factor $\sim 4$ and $\sim 10$ higher 
for \psr\ in the optical and X-rays, respectively (Table~8). For a 
comparison we show also in Table~8 the parameters for the $\sim 10$ times 
older Vela pulsar, which is much less luminous and a less efficient optical 
and X-ray emitter, but still capable of powering a weak and less extended  
PWN around it. The Vela pulsar
has also a peculiar optical spectrum in comparison with the Crab pulsar 
with a possible excess in the near-IR and a dip in the U band 
(Shibanov et al.~2003). Based on the available data it is not yet clear 
whether these spectral peculiarities in the pulsar optical emission do 
indicate a spectral evolution with pulsar age (\psr\ has a spin-down age
which is $\sim 400$ years higher than the Crab pulsar) or whether they
are connected to the pulsar optical efficiency, or just reflects specific 
parameters (e.g., viewing angle  and magnetic field geometry) of each pulsar. 
Detecting \psr\ in the near-IR would allow us to understand to which 
extent its spectrum is similar to that of the Vela pulsar and other 
middle-aged pulsars detected in the optical range. We also note that the 
ratio of optical to X-ray luminosity, weighted by the observed spectral 
ranges (cf. the last column in Table~8), is for \psr\ about half the value
for the Crab pulsar, but comparable with that of the Vela pulsar.  

\subsection{Multiwavelength spectrum of the PWN} 

%-----------------------figure15------------------------------
\begin{figure*}[thb]
\begin{center}
\includegraphics[width=100mm, clip, angle=270]{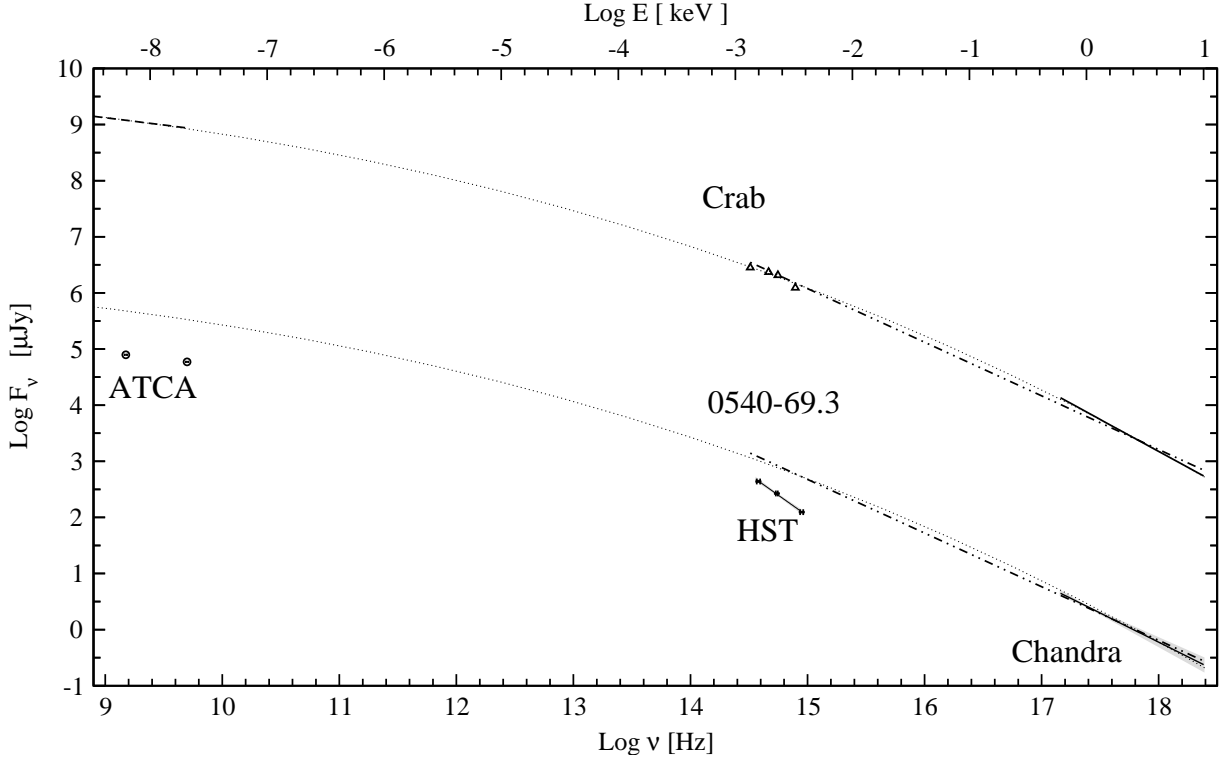}
\end{center}
\caption{Spatially averaged multiwavelength spectrum of the \psr\ PWN obtained
with the different telescopes indicated in the plot. The optical data are 
from the present work, whereas the radio and X-ray data are from Manchester 
et al. (1993b) and Kaaret et al. (2001), respectively. The filled regions 
(for the HST and Chandra data) show 1$\sigma$ uncertainty regions around 
the best-fit power laws indicated by thick lines. We have also included 
similar data for the Crab PWN, where the data are from Velusamy et al.
(1992), Veron-Cetty \& Woltjer (1993) and Kaaret et al. (2001) for
the radio, optical and X-ray bands, respectively. The dotted curved line is
a cubic spline fit for the multiwavelength spectrum of the Crab PWN, and 
the dashed-dotted line is a linear fit for the optical/X-ray region of the 
Crab PWN. These fits have been shifted to the level of the X-ray emission of 
the \psr\ PWN for a comparison. Note how the fits overshoot for \psr\ in both 
the optical and the radio.
} \label{f:pwnOptX}
\end{figure*}
%------------------------------------------------------------
A compilation of our optical data for the PWN around \psr\ together with 
radio and X-ray data is shown as an unabsorbed multiwavelength spectrum in 
Fig~\ref{f:pwnOptX}. The spectrum suggests the same double knee connection
between the optical and X-ray spectral parts as for the pulsar (cf. above). 
However, the assumed knee-breaks appear to be less pronounced than for the 
pulsar because the order of magnitude higher ratio for the intrinsic 
optical to X-ray flux for the PWN than for the pulsar. This is also
reflected in the PWN luminosities presented in Table~9, where we have also 
collected information on the spectral indices and luminosities of the Crab 
and Vela PWNs. The smoothness of the knee in the PWN spectrum as compared 
to the pulsar spectrum may be explained by propagation 
effects of the relativistic particles generated in the pulsar magnetosphere 
and moving through the PWN. If the particle energy distribution function 
contains features which are reflected in the pulsar spectrum, these features 
should become less pronounced due to such propagation effects.  
In Fig.~\ref{f:pwnOptX}, we also note a smaller difference between the 
radio and optical fluxes for the PWN, compared to for the pulsar, which 
may be also be caused by such propagation effects. 

Unfortunately, there is no data of similar quality on the spatially 
averaged optical spectrum of the Crab PWN, but spectra of specific 
signatures of the Crab PWN (knots, wisps etc.) are significantly steeper 
than the pulsar spectrum (Sollerman~2003).
On a larger scale, the continuum optical emission of the whole Crab Nebula,
at 10\arcsec\ spatial resolution (Veron-Cetty \& Woltjer 1993), shows a
similar spectral behavior. In particular, a bright area roughly
overlapping in position and morphology with the Crab X-ray PWN was resolved. 
Its spatially averaged spectrum is steeper than the spectrum of the 
pulsar. 
We see the same behavior for the whole 
PWN of \psr\ and for some parts of it (cf.~Figs.~6 and 7), whereas the
much fainter Vela PWN has not yet been detected in the optical 
(Mignani et al.~2003). However, some hints of specific structures in the Vela
case, clearly detected in X-rays (Helfand et al.~2001; Pavlov et al.~2001),
may have been detected also in the near-IR JH bands (Shibanov et al.~2003). 
There is no reliable spectral information on them and their identification 
still have to be confirmed by deeper observations both in the optical and IR. 

The optical efficiency of the PWN of \psr\ is a factor of $\sim 30$ higher 
than for the pulsar. This is markedly different from the situation 
in X-rays where the efficiency of the PWN is only $\sim 4$ times higher 
than for the pulsar (Kaaret et al.~2001).  
If we assume that the bright optical areas of Veron-Cetty \& Woltjer (1993) 
are associated with the  the optical emission of the whole Crab  PWN, 
we obtain a similar situation also for this pulsar with its $\sim 50$ 
higher optical efficiency of the PWN compared to that of the pulsar, while     
the efficiency ratio is only a factor of 2--3 in X-rays.  
This may give a hint on the particle 
energy distribution responsible for the synchrotron emission of the PWN. 

To estimate the ratio of the efficiencies in the optical and X-rays 
we normalized the luminosity in each domain to the respective energy range 
for which the luminosity was calculated. The ranges are 
3370 -- 7870 \AA\  (i.e., 1.58--3.68 eV) and 0.6 -- 10 keV. As seen in 
Tables~8 and 9, the normalized optical luminosity
(${L^{\rm O}\Delta E^{\rm X}~/~L^{\rm X}\Delta E^{\rm O}}$, where the
parameters are defined in Table 8) is $\sim 16$ and $\sim 140$ 
times higher than that in X-rays for the pulsar and its PWN, respectively.
The inferred values for the Crab are even higher 
($\sim35$ and  $\sim 870$, respectively). This is natural since 
the rotational loss of the Crab pulsar is $\sim 3$ times higher and its 
multiwavelength spectrum does not show a depression in the optical, 
as \psr\ appears to have. 
%-----------------------table9------------------------------
\begin{table*}[hbt] 

\caption{ Same as in Table~\ref{t:psr-lum} but for the PWNs of the same pulsars. 
Information on the PWN size and the ratios of the pulsar 
to PWN luminosities  in the optical and X-rays for each pulsar is included.}  
\label{t:pwn-lum}
\begin{center}
\begin{tabular}{lcccccccccc}
\hline\hline 
PWN & size & ${\rm \alpha^{O}_{\nu}}$ & ${L^{\rm O,a}}$ & ${\rm \eta ^{O}}$ & ${\alpha^{\rm X}_{\nu}}$ 
& ${L^{\rm X,b}}$ & ${\rm \eta ^{X}}$ & ${L^{\rm O}\Delta E^{\rm X} \over L^{\rm X}\Delta E^{\rm O}}$ 
& ${L^{\rm O}_{\rm psr}/L^{\rm O}_{\rm pwn}}$ & ${L^{\rm X}_{\rm psr}/L^{\rm X}_{\rm pwn}}$ \\  
    & pc & & ${\rm 10^{33}~erg~s^{-1}}$ & ${\rm 10^{-5}}$ &  & ${\rm 10^{36}~erg~s^{-1}}$ & ${\rm 10^{-3}}$ &  &  & \\
\hline  
Crab & 1.5 & 0.92$^c$ & 4240$^c$  & 920 & 1.14$^b$  & 21.8$^d$ & 47.5 &  867 & 0.0017 & 0.046  \widerul\\
0540  &  0.6--0.9  & 1.5$^e$ & 366$^e$ & 245 &  1.04$^b$  & 12$^d$  & 79.7 & 136 & 0.03 & 0.26  \widerul\\    
Vela  & 0.14 & -- & -- & -- & 0.5$^f$ & ${\rm 6.8^d\times 10^{-4}}$ & ${\rm 9.8\times 10^{-2}}$ & -- & -- & 0.34 \widerul\\   
\hline
\end{tabular}
\begin{tabular}{ll} 
$^a$~For the optical range 1.57--3.68 eV. & $^e$~This paper. \\
$^b$~For the X-ray range 0.6--10 keV.     & $^f$~From Pavlov et al.~(2001b); ${L^{\rm X}}$ rescaled to the 0.6--10 keV \\
$^c$~Optical fluxes are taken from Veron-Cety \& Woltjer (1993). &  range and $d=0.29$~kpc. \\
$^d$~Kaaret et al. (2001); $ {L^{\rm X}}$ rescaled to the 0.6--10 keV range. & \\
\end{tabular} 
\end{center}
\end{table*}

The PWN of \psr\ has similar sizes in the optical (this paper) and in X-rays  
(\cite{GW00}; \cite{Kaa01}) and extends up to 4\arcsec\ away from the 
pulsar, which corresponds to $\approx$1 pc at 51 kpc. However, the brighter
emission is confined to $0.6\times0.9$~pc (cf. Fig. 5). This is a factor 
of $\sim 2$ smaller than the size of the Crab PWN which has also similar 
sizes in the optical and in X-rays (\cite{Hester02}\footnote {see also Chandra 
public images 
at chandra.harvard.edu/photo/2002/0052/0052\_xray\_opt.jpg}). 
The smaller size of the \psr\ PWN is in rough agreement with the expected  
(e.g., Kennel \& Coroniti 1984) 
PWN size scaling with pulsar spindown 
luminosity, $\propto {\dot E}^{0.5}$. A similar proportionality has been 
found from the comparison of the Crab and Vela PWNs (Helfand et al. 2001; 
Pavlov et al.~2001b). 

The higher optical and X-ray efficiencies of \psr, as compared with the Crab
pulsar, are not reflected in a simple way in its PWN efficiencies. 
For instance, the X-ray efficiency of the \psr\ PWN is almost twice as high 
as that of the Crab PWN, while its optical efficiency is $\sim 4$ times lower 
than in the Crab case. The reason for this is unclear and may be explained 
either by the propagation effects discussed above, or by different pulsar 
environments, or a combination of these effects. The pulsar contribution to 
the total pulsar+nebula X-ray luminosity is 4.5\%, 21\%, and 25\% for 
the \psr, Crab, and Vela pulsars, respectively. The contribution is smaller 
in the optical and ranges from 0.17\% for the Crab to 3\% for \psr. 
The Vela PWN has not been detected in the optical range, but we note
that even for this much older and fainter pulsar, the PWN dominates      
the total X-ray luminosity, and that its contribution is comparable  
to that of \psr, although $\dot{E}$ is an order of the magnitude smaller. 
This cannot be explained by a simple scaling with the spindown luminosity, 
as in case of the PWN sizes, and requires additional studies.               

Our results do not reveal any significant variation 
of the spectral index over the torus and jet like structures 
of the PWN of \psr, although a marginal inverse correlation between 
the spectral hardness and brightness cannot be excluded 
at $1.4\sigma$ level (cf.~Fig.~10 and Sect. 2.5 for more details). 
No significant variation of the spectral index was also found in 
recent X-ray studies of the Crab PWN along its torus and the cores of the 
PWN jets (Mori et al. 2004). This suggests that for both these PWNs 
the energy spectra of the emitting electrons and positrons injected  
in these two different directions by the shocked pulsar wind 
are similar. This fact, as well as the similar sizes 
of the these parts of PWNs in the optical and X-rays suggest 
that the particle spectra in the bright, central parts of PWNs 
are not affected by synchrotron cooling. In the Crab case 
the spectral softening and hence the synchrotron losses 
become significant only in the faint outermost regions 
of the PWN detected in X-rays (Mori et al. 2004; Weisskopf et al.~2000) 
and in the UV/optical ranges (Scargle 1969; Veron-Cetty \& Woltjer~1993; 
Hennessy et al.~1992). A similar dependence may be seen 
in the Vela PWN (Pavlov et al.~2001b). This imposes important 
constraints on the pulsar wind models.   
Deeper observations in the optical and X-rays are needed
to study the faint outer regions of the PWN of \psr, and   
to test the reality and magnitude of the inverse hardness/brightness 
correlation. We note a significant surface brightness 
asymmetry with the respect to the pulsar position along the major 
axis of the torus structure of the PWN of \psr\ seen in both the 
optical and in X-rays. This asymmetry can hardly be explained within 
the framework of axisymmetrical pulsar wind models by simply 
invoking Doppler boosting and relativistic aberration effects, as has been
done for the near and far sides of the Crab PWN torus.  
The asymmetry is more likely to be caused by plasma instabilities  
in the internal parts of the pulsar wind flow, or by asymmetry of 
the SN ejecta. The latter is also discussed from another point of 
view in Sect. 3.3.

\subsection{Proper motion and ejecta asymmetry} 
In Sect. 2.6 we found the indicative result 
that \psr\ has a high proper motion corresponding
to 1190$\pm560\kms$, and that the direction is consistent with the pulsar 
moving along the southern jet axis. A pulsar velocity of $\sim 1000 \kms$ is
high, but not exceptional. In a recent VLA study of pulsar proper motions,
Brisken et al. (2003) show that there is a relatively large fraction of
pulsars which have velocities in excess of $500 \kms$. Peng et al. (2003) 
quantify this number to be $\sim 16\%$. 

Even if our results for the proper motion 
of \psr\  have only a $2\sigma$ significance,  
and a third epoch 
of HST imaging is needed to test whether or not  
\psr\ belongs to this  
high-velocity class of objects, we cannot avoid connecting the origin 
of the possibly large velocity of \psr\ to the origin of 
the significant redshift of the inner part of SNR~0540-69.3 which has been 
estimated to be several hundred$\kms$ (Kirshner et al. 1989; Serafimovich 
et al. 2004). The most straightforward interpretation of this is that the 
0540-69.3 system could be the result of a very asymmetric explosion.

Evidence of asymmetric supernova ejecta in core-collapse supernovae is 
abundant, and perhaps of specific interest for the 0540-69.3 system with
its asymmetric inner ejecta structure (e.g., Kirshner et al. 1989), is that 
in general the asymmetry appears to increase with depth into the ejecta 
(see Akiyama et al. 2003 and references therein). The prime example is 
SN 1987A where the powering by radioactive nucleids in the center occurs along 
bipolar jets that are likely to be aligned with the rotational axis of the 
presupernova (Wang et al. 2002). Assuming that also the explosion in 
core-collapse supernovae itself could be jet-induced,
Khoklov et al. (1999) find from 2D-modeling that pulsar kick 
velocities of $\sim 1000 \kms$ can be achieved. The pulsar would move along 
the jet axis, consistent with our tentative finding for \psr, but to 
reach a velocity as high as $1000 \kms$ a large difference in momentum 
between the two jets is required. As pointed out by Lai (2000), it is not 
clear what could give rise to such a difference.

Lai et al. (2001) discuss various models how to produce large pulsar kick 
velocities in the context of the aligned spin axis and pulsar proper motion
in Crab and Vela (cf. Sect. 2.6), and their discussion may now also apply
to \psr. The conclusion is that spin-kick alignment requires fast, perhaps 
close to break-up rotation, at the pulsar birth. There is 
observational evidence pointing in the same direction at least for the Crab 
pulsar (Atoyan 1999; Sollerman et al. 2001). 
We note that among the various models reviewed by Lai et al. (2001), the  
hydrodynamically driven kicks may face a problem in reproducing kick 
velocities of the order we infer for \psr, and Fryer (2004) 
finds that even in his most asymmetric 3D-models of an exploding $15 \Msun$ 
star, the neutrino emission becomes asymmetrically emitted, thereby damping 
out the hydrodynamical pulsar kick. Fryer suggests that one way to obtain fast 
pulsars is to rely on them being produced by low-mass progenitors 
($8-12 \Msun$). This is, however, not a likely explanation for the 
0540-69.3 system, which most probably originates from a $\sim 20 \Msun$ 
progenitor. The recent models of Scheck et al. (2004) appear to be more 
successful in producing pulsars with high kick velocities.  
It could also be that models including rotation
will alter the results of the hydrodynamically driven kick models (cf.
Lai et al. 2001), perhaps providing a link to the results of
Khoklov et al. (1999). Further observations of the 0540-69.3 system will
show if it can add to Crab and Vela as a testbed for different kick scenarios.

\begin{acknowledgements}
We are grateful to Jelle de Plaa for sharing with us his reduced ROSAT
and RXTE data of \psr,  to Lucien Kuiper for high energy 
data of the Crab pulsar, and to the referee, Patricia Caraveo, for useful 
comments allowing us to improve several important points in the text. 
Partial support for this work was provided by RFBR 
(grants 02-02-17668, 03-02-17423 and 03-07-90200).
and support was also given by The Royal Swedish Academy of Sciences.
The research of PL is further sponsored by the Swedish Research Council.
PL is a Research Fellow at the Royal Swedish Academy supported by a grant 
from the Wallenberg Foundation. The work was initialized while NIS was 
supported by a stipend from The Swedish Institute.
\end{acknowledgements}
{}

\end{document}